\documentclass[12pt]{article}
\usepackage{amsmath}
\usepackage{graphicx,subcaption}
\usepackage{enumerate}
\usepackage{natbib}
\usepackage{url} 
\usepackage{color,comment}
\usepackage[]{changes}

\RequirePackage[OT1]{fontenc}
\RequirePackage{amsthm,amsmath,bm,hyperref}
\newcommand{\pkg}[1]{{\normalfont\fontseries{b}\selectfont #1}}
\let\proglang=\textsf
\usepackage{graphicx}
\usepackage{float}
\usepackage{amsthm}
\usepackage{amsfonts}
\usepackage{amsmath,mathtools}
\DeclareMathAlphabet\mathbfcal{OMS}{cmsy}{b}{n} 
\usepackage{multirow}
\newtheorem{theorem}{Theorem}[section]
\newtheorem{lemma} {Lemma}[section]
\newtheorem{proposition} {Proposition}[section]

\newcommand{\blind}{1}

\begin{document}

\def\spacingset#1{\renewcommand{\baselinestretch}%
{#1}\small\normalsize} \spacingset{1}


\if1\blind
{
  \title{\bf Functional Singular Spectrum Analysis}
  \author{Hossein Haghbin \\
    Department of Statistics, Persian Gulf University, Iran\\
    and \\
    Seyed Morteza Najibi \\
    Department of Statistics, Shiraz University, Iran\\
    and \\
    Rahim Mahmoudvand \\
    Department of Statistics, Bu-Ali Sina University, Iran\\
    and \\
    Jordan Trinka, and Mehdi Maadooliat \\
    Department of MSSC, Marquette University, USA}
    \date{}
  \maketitle
} \fi

\if0\blind
{
  \bigskip
  \bigskip
  \bigskip
  \begin{center}
    {\LARGE\bf Functional Singular Spectrum Analysis}
\end{center}
  \medskip
} \fi

\bigskip
\begin{abstract}
In this paper, we introduce a new extension of the Singular Spectrum Analysis (SSA) called functional SSA to analyze functional time series. The new methodology is developed by integrating ideas from functional data analysis and univariate SSA. We explore the advantages of the functional SSA in terms of simulation results and two real data applications. We compare the proposed approach with Multivariate SSA (MSSA) and dynamic Functional Principal Component Analysis (dFPCA). The results suggest that further improvement to MSSA is possible, and the new method provides an attractive alternative to the dFPCA approach that is used for analyzing correlated functions. We implement the proposed technique to an application of remote sensing data and a call center dataset. We have also developed an efficient and user-friendly R package and a shiny web application to allow interactive exploration of the results. 
\end{abstract}

\noindent%
\vfill

\newpage
\spacingset{1.5} 
\section{Introduction} \label{sec:int}
One of the popular approaches in the decomposition of time series is accomplished using the rates of change. In this approach, the observed time series is partitioned (decomposed) into informative trends plus potential seasonal (cyclical) and noise (irregular) components. Aligned with this principle, Singular Spectrum Analysis (SSA) is a model-free procedure that is commonly used as a nonparametric technique in analyzing the time series. SSA is intrinsically motivated as an exploratory and model building tool rather than a confirmatory procedure \citep{golyandina2001analysis}. SSA does not require restrictive assumptions such as stationarity, linearity, and normality. It can be used for a wide range of purposes such as trend and periodic component detection and extraction, smoothing, forecasting, change-point detection, gap filling, causality and so on; \citep[see, e.g. ][]{golyandina2001analysis, moskvina2003algorithm, kondrashov2010gap, golyandina2007caterpillar, mohammad2011comparing, mahmoudvand2016missing, rodrigues2016correlation}. 

The implementation of SSA over time series is similar to that of Principal Components Analysis (PCA) of multivariate data. In contrast to PCA, which is applied to a data matrix with independent rows, SSA is applied to a time series. It provides a representation of the given time series in terms of eigentriples of a so-called trajectory matrix \citep{alexandrov2008method}. 

Up to this day, many studies have been published with extensions and applications of SSA. Extensions to a multivariate model as well as to a two-dimensional setting can be found, e.g., in \cite{golyandina2013singular, golyandina2018singular, hassani2018singular} and references therein. In the regular SSA, we assume that the observation at each time point is scalar, vector or array. As a matter of interest, one may consider a series of curves observed over time, and use the basics of Hilbert space in the functional data analysis (FDA) framework to introduce the concept of functional SSA (FSSA).
 
While the research in FDA has grown extensively in recent years, there have been relatively few contributions dealing with functional time series (FTS); see, e.g., \cite{hormann2012} and \cite{bosq2000}. Although most of the current FTS approaches focus on a parametric fit for inferences and forecasting, there exist some other approaches in the literature that extend FPCA to incorporate the temporal correlation of FTS. For instance, \citet{hormann2015} introduced dynamic FPCA (dFPCA) to analyze FTS. This approach assumes the strong assumption of stationarity which is not generally held in practice.
It would be of interest to non-parametrically decompose a nonstationary FTS to reveal the respective trends plus seasonal and irregular components in an appropriate manner. Consistent with this approach, and as a first step, \citet{fraiman2014detecting} introduced a new concept of trends for the FTS. Furthermore, \citet{hormann2018testing} considered the periodic components for the FTS and derived several procedures to test the periodicity using frequency domain analysis. To the best of our knowledge, existing studies mainly focus on detecting rather than extracting interpretable components. 

Since one of the primary missions of SSA is to extract trends and periodic components of a regular (non-functional) time series, it would be rational to establish a similar elegant nonparametric procedure to extract such components in FTS. In this paper we use the basics of SSA and multivariate functional PCA (MFPCA), introduced in \cite{happ2016,chiou2014}, to develop FSSA. In a nutshell, the core of SSA is to use PCA on the variables being lagged versions of a single time series. Since a lagged vectors of FTS forms a multivariate functional variable, we use the theory of MFPCA to develop the FSSA procedure. The new methodology, FSSA, not only can serve as a nonparametric dimension reduction tool to decompose the functional time series; it can also be used as a visualization tool to illustrate the concept of seasonality and periodicity in the functional space over time.

\begin{figure}[!h]
\begin{center}
    \begin{subfigure}[b]{0.4\textwidth}
	\includegraphics[page=1,width=\textwidth]{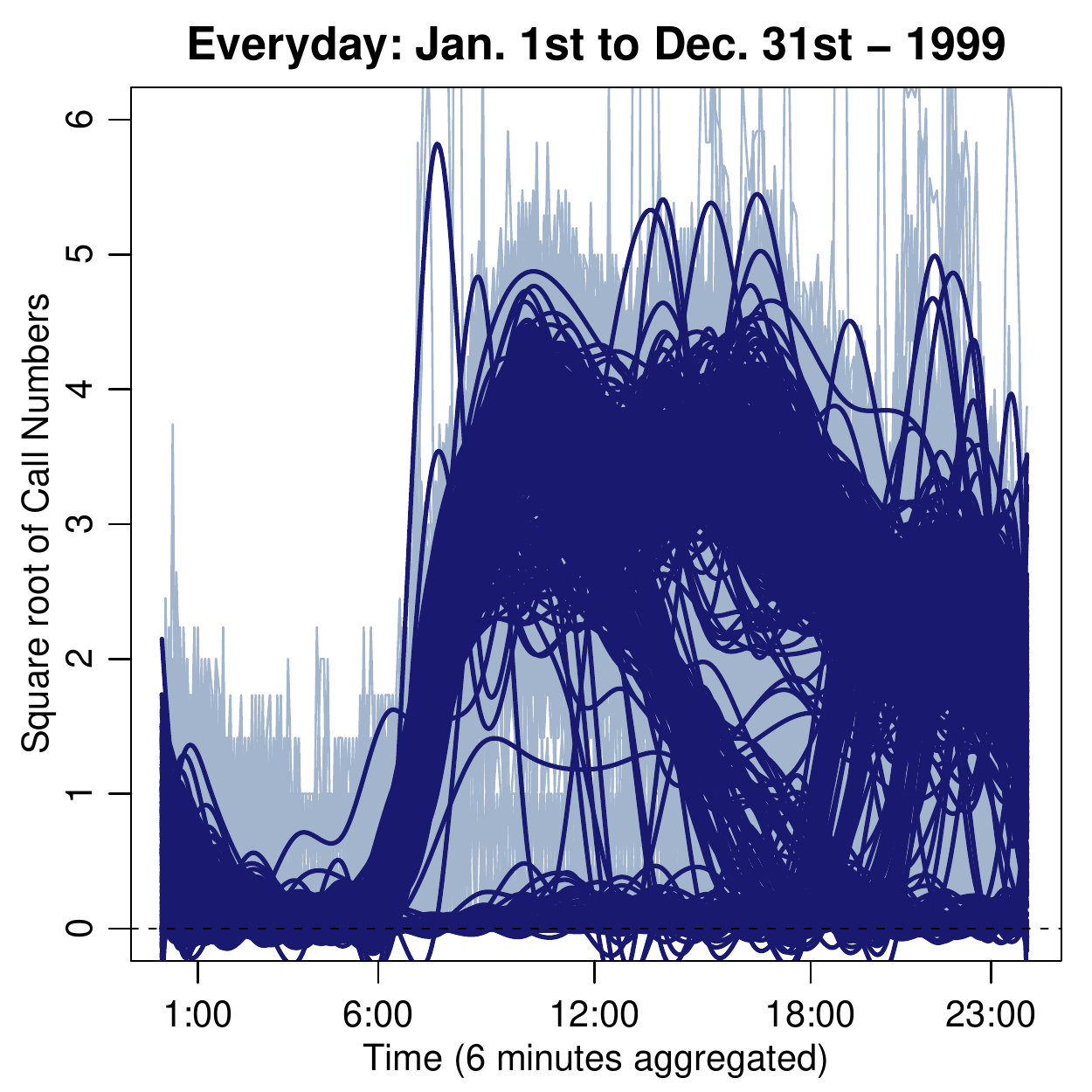}
    \end{subfigure}
    \begin{subfigure}[b]{0.4\textwidth}
	\includegraphics[page=2,width=\textwidth]{Figure1}
    \end{subfigure}
\caption{The number of calls to a call center between January 1\textit{st} to December 31\textit{st} in the year 1999.}
\label{fig:motivating_call}
\end{center}
\end{figure}

\begin{figure}[!h]
\begin{center}
\begin{subfigure}[b]{0.24\textwidth}\includegraphics[page=1,width=\textwidth]{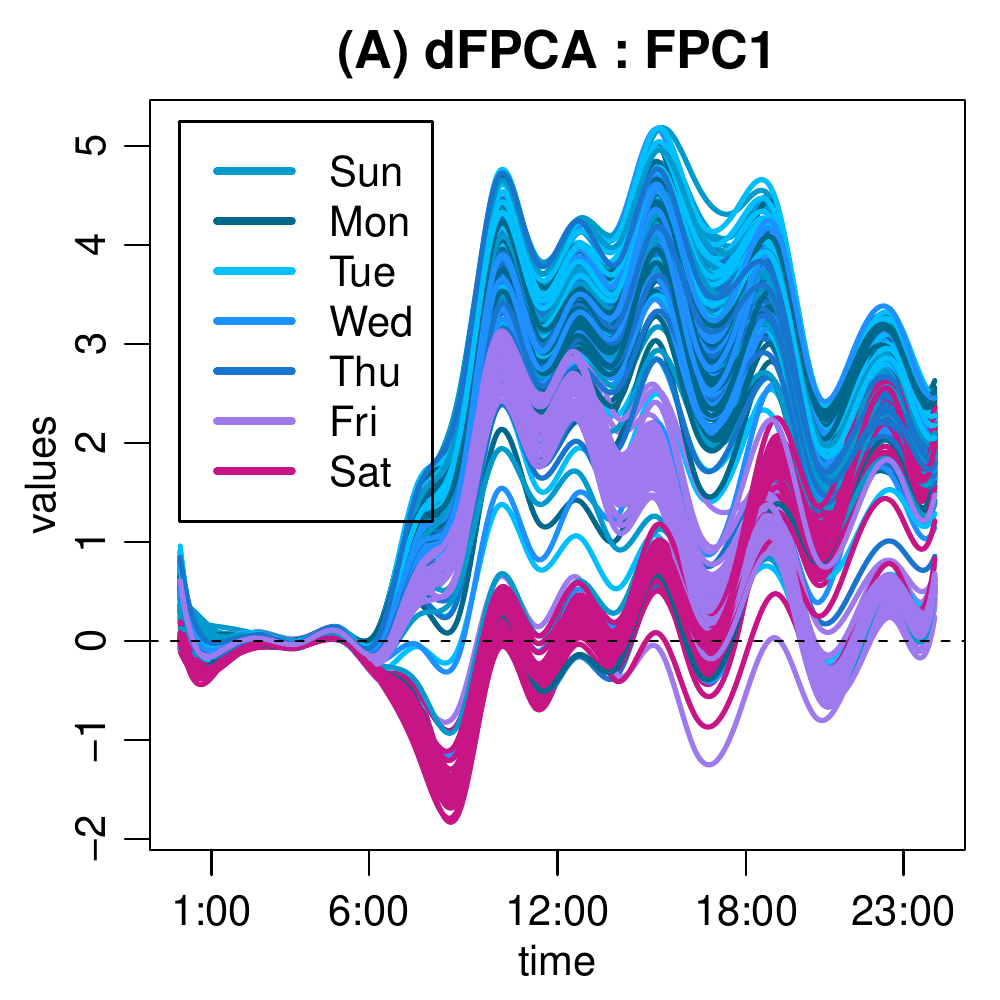}\end{subfigure}
\begin{subfigure}[b]{0.24\textwidth}\includegraphics[page=2,width=\textwidth]{Figure2-dFPCA}\end{subfigure}
\begin{subfigure}[b]{0.24\textwidth}\includegraphics[page=3,width=\textwidth]{Figure2-dFPCA}\end{subfigure}
\begin{subfigure}[b]{0.24\textwidth}\includegraphics[page=4,width=\textwidth]{Figure2-dFPCA}\end{subfigure}
\begin{subfigure}[b]{0.24\textwidth}\includegraphics[page=1,width=\textwidth]{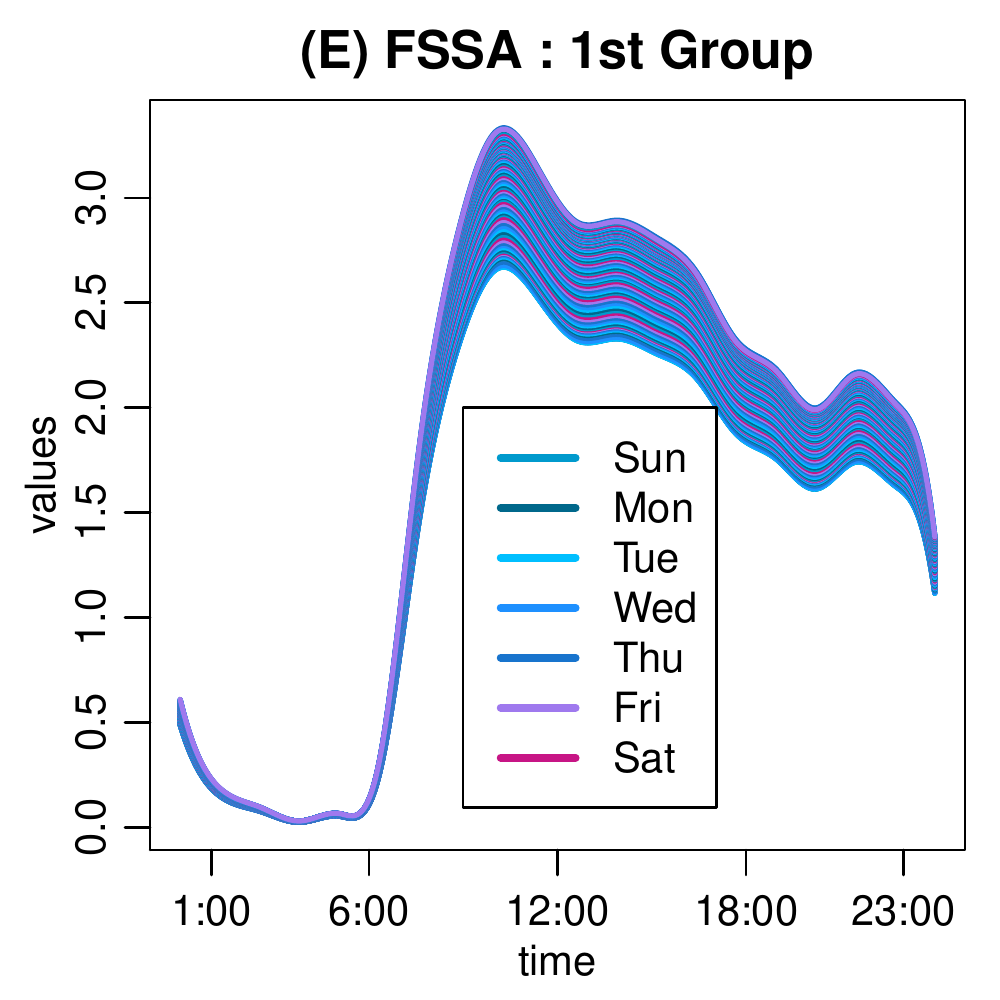}\end{subfigure}
\begin{subfigure}[b]{0.24\textwidth}\includegraphics[page=2,width=\textwidth]{Figure2-FSSA}\end{subfigure}
\begin{subfigure}[b]{0.24\textwidth}\includegraphics[page=3,width=\textwidth]{Figure2-FSSA}\end{subfigure}
\begin{subfigure}[b]{0.24\textwidth}\includegraphics[page=4,width=\textwidth]{Figure2-FSSA}\end{subfigure}
\caption{Four leading DFPCA FPC's associated to the call center data (top) and FSSA reconstructed series after grouping (bottom).}
\label{fig:call_recons}
\end{center}
\end{figure}

In order to depict the idea of our approach and to show its utility, consider the following motivating example involving a real dataset which is described in detail in the supplementary materials. This data provides the intraday number of calls to a call center, during different times of the days for one year. The associated $365$ curves is represented in an overlapping pattern in Figure \ref{fig:motivating_call} (left). In Figure \ref{fig:motivating_call} (right) we investigate the pattern among weekdays and weekend days. As we can see, the intraday patterns of weekends (Friday and Saturday) are significantly different from workdays while workdays seem to have similar patterns. Investigators used variants of FPCA to analyze the call center data in literature \citep{shen2005analysis, huang2008functional, maadooliat2015integrating}. For illustration purpose, we compare the results of the proposed method (FSSA) and dFPCA on this dataset. Figure \ref{fig:call_recons} (top) presents the projection of the data into the first four FPCs of the dFPCA obtained from the \pkg{freqdom.fda} package in \proglang{R} \citep{hormann2015}. We used seven different colors to differentiate between different days of a week. As one may observe, visually, there is no clear separation in either one of the FPC graphs in the top row. Although this may not be surprising, as the purpose of PCA of any type is to reduce the dimensionality, and not necessary decompose the data into regular trends, periodic and irregular components. In contrast, the grouping results that we obtained using the FSSA on the call center data are given in Figure \ref{fig:call_recons} (bottom). It can be seen that the functional behavior of seven days of a week can be well-distinguished, visually, using either one of the last two groups (groups 3 and 4).

The rest of the paper is organized as follows. Section~\ref{ssa-method} reviews the core of SSA for completeness. Section ~\ref{fssa-method} presents the theoretical foundations and some properties of the proposed method (FSSA), and Section~\ref{sec:Practical Imp.} provides implementation details. Section~\ref{simul} reports simulation results to illustrate the use of the proposed approach in analyzing FTS, and to compare it with MSSA and dFPCA. Application to a real data example on remote sensing is given in Section~\ref{sec:NDVI}. Section~\ref{discussion} provides some discussions and concluding remarks.
 
\section{General scheme of SSA}\label{ssa-method}
As we mentioned in Section \ref{sec:int}, SSA can be used for many purposes. However, as we intend to introduce the functional version of SSA for decomposing FTS, we review a general scheme of SSA to perform time series decomposition in this section. 

\subsection{Univariate SSA}
Throughout this section, we consider $y_i$'s are elements of Euclidean space $\mathbb{R}$. Suppose that ${\bf y}_N = (y_1, y_2, \ldots , y_N)^\top$ is a realization of size $N$ from a time series. The basic SSA algorithm consists of four steps: Embedding, Decomposition, Grouping, and Reconstruction.

\subsubsection*{Step 1. Embedding}
This step generates a multivariate object by tracking a moving window of size $L$ over the original time series, where $L$ is called \textit{window length} parameter and $1<L<N$. Embedding can be regarded as a mapping operator $\mathcal{T}$ that transfers the series ${\bf y}_N$ into a so-called \textit{trajectory} matrix ${\bf X}$ of dimension $L\times K$, defines by
	\begin{equation}\label{traj}
	{\bf X}=\mathcal{T}\left({\bf y}_N\right)=\left[{\pmb x}_1, \ldots, {\pmb x}_K \right],
	\end{equation}
where $K=N-L +1$ and ${\pmb x}_j=(y_j, y_{j+1}, \ldots, y_{j+L-1})^\top$, for $j=1,\ldots, K$, are called \textit{lagged vectors}. Note that the trajectory matrix ${\bf X}$, is a \textit{Hankel} matrix, which means that all the elements along the anti-diagonals are equal. Indeed, the embedding operator $\mathcal{T}$ is a one-to-one mapping from $\mathbb{R}^N$ into $\mathbb{R}^{L\times K}_H\subseteq\mathbb{R}^{L\times K}$, where $\mathbb{R}^{L\times K}_H$ is the set of all $L\times K$ Hankel matrices. 

\subsubsection*{Step 2. Decomposition}
In this step, the singular value decomposition (SVD) for the trajectory matrix is computed as: 
	\begin{equation}\label{svd}
	{\bf X}=\sum_{i=1}^L {\sqrt{\lambda_i}}{\bf u}_i{\bf v}_i^\top=\sum_{i=1}^L {\bf X}_i.
	\end{equation}
where $\sqrt{\lambda_i}$ is the $i^{th}$ singular value of $\bf X,\ {\bf u}_i$ and ${\bf v}_i\ $ are the associated (orthonormal) left and right singular vectors, and ${\bf X}_i=\sqrt{\lambda_i}{\bf u}_i{\bf v}_i^\top$ is called the respective \textit{elementary} matrix. Note that ${\bf u}_i$ is an eigenvector of ${\bf XX}^\top$ corresponding to the eigenvalue $\lambda_i$. Moreover, it yields that
	\begin{equation}\label{eq: primary}
	{\bf X}_i={\bf u}_i{\bf u}_i^\top{\bf X}=\left[ ({\bf u}_i\otimes{\bf u}_i) {\bf x}_1, \ldots, ({\bf u}_i\otimes{\bf u}_i ){\bf x}_K\right],
	\end{equation}
where, in this section, $\otimes$ denotes the outer (tensor) product of two vectors.

\subsubsection*{Step 3. Grouping}
Consider a partition of the set of indices $\{1, 2, \ldots, r\}$, where $r$ is the rank of the matrix $\bf{X}$, into $m$ disjoint subsets $\{{I_1}, {I_2}, \ldots, {I_m}\}$. For any positive integer $q$, i.e. $1\leq q \leq m$, the matrix ${\bf X}_{I_q}$ is defined as ${\bf X}_{I_q}=\sum_{i\in {I_q}}{\bf X}_i$. Thus, by the expansion \eqref{svd} we have the \textit{grouped matrix decomposition }
	\begin{equation}\label{svd2}
	{\bf X}={\bf X}_{I_1}+{\bf X}_{I_2}+\cdots+{\bf X}_{I_m}.
	\end{equation}
Each group in \eqref{svd2} should correspond to a component in time series decomposition. These components can be considered as trend, cycle, seasonal, noise, etc.

\subsubsection*{Step 4 Reconstruction}
Finally, the resulting matrices ${\bf X}_{I_q}$ in \eqref{svd2}, are transformed back into the form of the original series ${\bf y}_N$ by an inverse operator $\mathcal{T}^{-1}$. In order to do this, first, it is necessary that each matrix ${\bf X}_{I_q}$ to be approximated by a matrix in $\mathbb{R}^{L\times K}_H$. This approximation is performed optimally in the sense of orthogonal projection of ${\bf X}_{I_q}$ on $\mathbb{R}^{L\times K}_H$ with respect to the \textit{Frobenius} norm. Denote this projection by $\Pi:\mathbb{R}^{L\times K}\rightarrow \mathbb{R}^{L\times K}_H$. It is shown that the projection $\Pi$ is the averaging of the matrix elements over the antidiagonals $i + j = const$. By combining the results of this step and \eqref{svd2}, we obtain the final decomposition of the series in the form of 
	\begin{equation}\label{ssa}
	{\bf y}_N=\tilde{\bf y}_1+\tilde{\bf y}_2+\ldots+\tilde{\bf y}_m,
	\end{equation}
where $\tilde{\bf y}_q=\mathcal{T}^{-1}\Pi({\bf X}_{I_q})$, for $q=1,\ldots, m$. 

The above algorithm can be extended to perform Multivariate SSA (MSSA) for analyzing multivariate time series. The only difference is in defining the trajectory matrix which can be defined by stacking univariate trajectory matrices horizontally or vertically \citep{hassani2018singular}.

It is well known that SSA does not require restrictive assumptions; however, it is ideal to have a time series with separable components. Therefore, we present tools to measure the separability of components in the next subsection. 

\subsection{Separability}
Let $\textbf{y}_N^{(i)}=\left\lbrace y_1^{(i)},\ldots, y_N^{(i)} \right\rbrace$, for $i=1,2$ be two time series and consider an additive model as $\textbf{y}_N=\textbf{y}_N^{(1)}+\textbf{y}_N^{(2)}$. The series $\textbf{y}_N^{(1)}$ and $\textbf{y}_N^{(2)}$ are called separable when each lagged vector of $\textbf{y}_N^{(1)}$ is orthogonal to the lagged vectors of $\textbf{y}_N^{(2)}$. To measure the degree of separability between two time series $\textbf{y}_N^{(1)}$ and $\textbf{y}_N^{(2)}$, \citet{golyandina2001analysis} introduced the so-called w-correlation
	\begin{equation}\label{wcor}
	\rho^{(w)}(\textbf{y}_N^{(1)},\textbf{y}_N^{(2)})=\dfrac{\left\langle \textbf{y}_N^{(1)} , \textbf{y}_N^{(2)} \right\rangle_w}
	{\sqrt{\left\langle \textbf{y}_N^{(1)} , \textbf{y}_N^{(1)} \right\rangle_w}\sqrt{\left\langle \textbf{y}_N^{(2)} , \textbf{y}_N^{(2)} \right\rangle_w}},
	\end{equation}
where, $\left\langle \textbf{y}_N^{(t)} , \textbf{y}_N^{(s)} \right\rangle_w=\sum_{i=1}^N w_i y_i^{(t)} y_i^{(s)}$ for $t,s=1,2$ and $w_i=\min\{i,L,N - i +1\}$. We call two series $\textbf{y}_N^{(1)}$ and $\textbf{y}_N^{(2)}$ w-orthogonal if $\rho^{(w)}(\textbf{y}_N^{(1)},\textbf{y}_N^{(2)}) = 0$ for approriate values of $L$ (see the next subsection for more details). Note that $\tilde{\bf y}_q$, $q=1,\ldots, m$, is the reconstructed component produced by the group $I_q$, and the matrix of ${\bm \rho}^{(w)} = \bigl\{\rho^{(w)}(\tilde{\bf y}_i,\tilde{\bf y}_j)\bigr\}_{i,j=1}^m$ is called w-correlation matrix. 

\subsection{Parameter Selection}
There are two basic parameters in SSA procedure; window length ($L$) and grouping parameters. Choosing improper values for these parameters yields an incomplete reconstruction and misleading results in subsequent analysis. In spite of the importance of choosing $L$ and grouping parameters for SSA, no ideal solution has been yet proposed. A thorough review of the problem shows that the optimal choice of the parameters depends on the intrinsic structure of the data and the purposes of the study \citep{golyandina2001analysis,golyandina2013singular}. However, there are several recommendations and rules that work well for a wide range of scenarios. It is recommended to select the window length parameter, $L$, to be a large integer that is multiple of the periodicities of the time series, but not larger than $\frac{N}{2}$. 

In addition, there are several utilities for effective grouping. These tools include analyzing the periodogram, paired plot of the singular vectors, scree plot of the singular values, and w-correlation plot; see \cite{golyandina2001analysis} for more details.

\section{Theoretical Foundations of FSSA}\label{fssa-method}
We start this section with the mathematical foundations that are used to develop the functional SSA procedure. From hereafter, we consider $\textbf{y}_N=\left( y_1,\ldots, y_N \right)^\top$ is a FTS of length $N$. This means that each elements $y_i:[0,1]\rightarrow \mathbb{R}$ belongs to $\mathbb{H}:=\mathcal{L}^2([0,1])$, the space of square integrable real functions defined on the interval $[0, 1]$. Here, the space $\mathbb{H}$ is a Hilbert space, equipped with inner product $\left\langle x,y\right\rangle=\int_0^1 x(s)y(s)ds$. For a given positive integer $k$, the space $\mathbb{H}^k$ denotes the Cartesian product of $k$ copies of $\mathbb{H}$; i.e. for an element ${\pmb x}\in \mathbb{H}^k$, it has the form ${\pmb x}(s)= \begin{pmatrix} x_1(s), x_2(s), \ldots, x_k(s) \end{pmatrix}^\top$, where $x_i\in \mathbb{H}$, and $s\in[0,1]$. Then $\mathbb{H}^k$ is a Hilbert space equipped with the inner product $\langle \pmb x,\pmb y\rangle_{\mathbb{H}^k}=\sum_{i=1}^k \langle x_i,y_i\rangle$. The norms will be denoted by $\Vert \cdot \Vert$ and $\Vert \cdot \Vert_{\mathbb{H}^k}$ in the spaces $\mathbb{H}$ and $\mathbb{H}^k$, respectively. For $x\in \mathbb{H}_1$, and $y\in \mathbb{H}_2$, where $\mathbb{H}_1$ and $\mathbb{H}_2$ are  two Hilbert spaces, we define the tensor(outer) product corresponding to the operator $x\otimes y: \mathbb{H}_1\rightarrow \mathbb{H}_2$, as ($x\otimes y)h:=\langle x, h \rangle y$, where $h\in \mathbb{H}_1$.

For positive integers $L\ \text{and}\ K$, we denote $\mathbb{H}^{L\times K}$ as the linear space spanned by operators $\mathbfcal{Z}: \mathbb{R}^K \rightarrow \mathbb{H}^L$, specified by $[{z}_{i,j}]_{i=1,\ldots,L}^{ j=1,\ldots,K}$ where
	\begin{equation}\label{eq: z operator}
	\mathbfcal{Z}\pmb{a}=
	\begin{pmatrix}	\sum_{j=1}^K a_j{z}_{1,j}\\ \vdots\\ \sum_{j=1}^K a_j{z}_{L,j} \end{pmatrix},
	\ {z}_{i,j}\in\mathbb{H},\ \text{and}\ \pmb{a}=(a_1,\ldots, a_K)\in\mathbb{R}^K.
	\end{equation}
We call an operator $\tilde{\mathbfcal{Z}}=[\tilde{{z}}_{i,j}]\in\mathbb{H}^{L\times K}$ Hankel if $\Vert\tilde{{z}}_{i,j}-g_s\Vert=0$, for some $g_s\in\mathbb{H}$, where $s=i+j$. The space of such Hankel operators will be denoted $\mathbb{H}_H^{L\times K}$. For two given operators $\mathbfcal{Z}_1=[\textbf{z}^{(1)}_{i,j}]$ and $\mathbfcal{Z}_2=[\textbf{z}^{(2)}_{i,j}]$ in $\mathbb{H}^{L\times K}$, define 
	\begin{equation*}
	\langle\mathbfcal{Z}_1,\mathbfcal{Z}_2\rangle_\mathcal{F}:=\sum_{i=1}^L\sum_{j=1}^K \langle{z}_{i,j}^{(1)},{z}_{i,j}^{(2)}\rangle.
	\end{equation*}
It follows immediately that $\langle \cdot , \cdot \rangle_\mathcal{F}$, defines an inner product on $\mathbb{H}^{L\times K}$. We will call it Frobenius inner product of two operators in $\mathbb{H}^{L\times K}$. The associated Frobenius norm is $\Vert \mathbfcal{Z} \Vert_\mathcal{F}=\sqrt{\langle\mathbfcal{Z},\mathbfcal{Z}\rangle_\mathcal{F}}$. Before discussing the FSSA algorithm, here we present a lemma that will be used in the last step of the proposed algorithm. Note that the Proofs for all lemmas, theorems and propositions are given in the supplementary materials.

\begin{lemma} \label{lemma:lse}
Let $x_i,\ i=1,\ldots, N$ be elements of the Hilbert space $\mathbb{H}$. If $\bar{x}=\frac{1}{N}\sum_{i=1}^N x_i$, then
	\begin{equation*}
	\sum_{i=1}^N\Vert x_i-\bar{x}\Vert^2 \leq \sum_{i=1}^N\Vert x_i-y \Vert^2,
	\end{equation*}
for all $y\in \mathbb{H}$.
\end{lemma}

\subsection{FSSA algorithm}
For an integer $1\leq L\leq{N}/{2}$, let $K=N-L+1$ and define a set of multivariate functional vectors in $\mathbb{H}^L$ by
	\begin{equation}\label{flvec}
	{\pmb x}_j(s):= \begin{pmatrix} y_j(s), y_{j+1}(s), \ldots, y_{j+L-1}(s)\end{pmatrix}^\top,\ \ j=1,\ldots, K,
	\end{equation}
where ${\pmb x}_j$'s denote the functional $L-$lagged vectors. The following algorithm provides the FSSA results in four steps. 

\subsubsection*{Step 1. Embedding}
Define the operator $\mathbfcal{X}:\mathbb{R}^K \rightarrow \mathbb{H}^L$ with
	\begin{equation}
\label{eq:traj}
	\mathbfcal{X}{\pmb a}:=\sum_{j=1}^K a_j{\pmb x}_j=
	\begin{pmatrix} \sum_{j=1}^K a_jy_j\\ \sum_{j=1}^K a_j y_{j+1}\\ \vdots\\ \sum_{j=1}^K a_j y_{j+L-1} \end{pmatrix},
	\ {\pmb a}=\left(a_1,\ldots, a_K\right)^\top \in\mathbb{R}^K.
	\end{equation}
We call $\mathbfcal{X}$ the \textit{trajectory operator.} It is easy to see that $\mathbfcal{X}=\mathcal{T}\textbf{y}_N$, where $\mathcal{T}$ is an operator from $\mathbb{H}^N$ to $\mathbb{H}_H^{L\times K}$. Evaluating $\mathbfcal{X} \pmb{a}$ at a given point $s\in [0,1]$ is same as the matrix product ${\bf X}(s)\pmb{a}$, where ${\bf X}(s)$ is an $L \times K$ Hankel matrix given by
	\begin{equation}\label{ftraj}
	{\bf X}(s)=\begin{bmatrix} {\pmb x}_1(s), \ldots, {\pmb x}_K(s) \end{bmatrix}.
	\end{equation}
	
\begin{proposition}\label{prop:trajO}
The operator $\mathbfcal{X}$ is a bounded linear operator. If we define $\mathbfcal{X}^*:\mathbb{H}^L \rightarrow \mathbb{R}^K$, given by
	\begin{equation}
	\mathbfcal{X}^*{\pmb z}=
	\begin{pmatrix} \sum_{i=1}^L \langle y_i, z_i\rangle\\ \sum_{i=1}^L \langle y_{i+1}, z_i\rangle\\ \vdots\\ \sum_{i=1}^L \langle y_{i+K-1}, z_i\rangle \end{pmatrix},
	\ {\pmb z}=\left(z_1,\ldots, z_L\right)^\top\in\mathbb{H}^L,
	\end{equation}
then $\mathbfcal{X}^*$ is an adjoint operator for $\mathbfcal{X}$.
\end{proposition}

\subsubsection*{Step 2. Decomposition}
Define the operator $\boldsymbol{\mathcal{S}}: \mathbb{H}^L\rightarrow \mathbb{H}^L $ by $\boldsymbol{\mathcal{S}}:=\mathbfcal{X}\mathbfcal{X}^*$. Therefore, for given ${\pmb z}\in \mathbb{H}^{L}$ it implies that
	\begin{align}\label{eq: s-operator}
	\mathbfcal{S}{\pmb z}&
	=\sum_{j=1}^K\sum_{i=1}^L \langle y_{i+j-1} , z_i \rangle {\pmb x}_j
	=\sum_{j=1}^K \langle {\pmb x}_j , {\pmb z} \rangle_{\mathbb{H}^L} {\pmb x}_j
	=\sum_{j=1}^K ({\pmb x}_j \otimes {\pmb x}_j) {\pmb z}.
	\end{align}

Here, the operator $\boldsymbol{\mathcal{S}}$ can be also considered as an $L \times L$ matrix with the operator entries $\mathcal{S}_{i,j}:\mathbb{H}\rightarrow \mathbb{H}$, given by $\mathcal{S}_{i,j}=\sum_{l=1}^K y_{i+l-1}\otimes y_{j+l-1}$, where $i,j=1, \ldots, L$. Note that, the operator $\mathcal{S}_{i,j}$ defines an integral operator on $\mathbb{H}$, associated to the kernel
	\begin{equation}\label{kernel1}
	{c}_{i,j}(s,u):=\sum_{k=1}^K y_{i+k-1}(s)y_{j+k-1}(u),\ \text{for}\ s,u\in[0,1].
	\end{equation} 
Let us define ${\bf C}:[0,1]\times[0,1]\rightarrow \mathbb{R}^{L\times L}$ to be a kernel matrix with the elements $\{c_{i,j}\}$. Note that ${\bf C}(s,u)={\bf X}(s){\bf X}(u)^\top$. It is easy to show that the associated integral operator of ${\bf C}$ is $\mathbfcal{S}$, i.e,
	\begin{align}\label{s-oper}
	{\mathbfcal{S}}{\pmb z}(u)&= \int_0^1 {\bf C}(s,u) {\pmb z}(s)ds=
	\begin{pmatrix} \sum_{i=1}^L\int_0^1 c_{i,1}(s,u) z_i(s) ds\\ \vdots \\ \sum_{i=1}^L\int_0^1 c_{i,L}(s,u) z_i(s) ds \end{pmatrix}.
	\end{align}

\begin{proposition}\label{prop_compact}
The operator $\boldsymbol{\mathcal{S}}$ defined in \eqref{eq: s-operator} is a linear, self-adjoint, positive definite, bounded, continuous and compact operator. 
\end{proposition}

By the results of the Proposition \ref{prop_compact} and the Hilbert-Schmidt Theorem (e.g. \cite{simon1980methods}, Thm. VI.16), it follows that there exists an orthonormal basis system $\left\lbrace \pmb{\psi}_i, \ i\in \mathbb{N}\right\rbrace$ of $\mathbb{H}^L$ such that 
\begin{equation}
\label{eq:eigen}
	\mathbfcal{S} \pmb{\psi}_i=\lambda_i \pmb{\psi}_i,\quad \text{and}\ \lambda_i \longrightarrow 0 \ \text{as}\ i \longrightarrow \infty.
	\end{equation} 
Furthermore, using the Spectral Theorem (e.g. \cite{werner2006funktionalanalysis}, Thm. VI.3.2.) implies
	\begin{equation}
	\mathbfcal{S} =\sum_{i=1}^\infty \lambda_i \pmb{\psi}_i\otimes \pmb{\psi}_i.
	\end{equation}
Since the kernel $\textbf{C}(s,u)$ is continuous, it admits the expansion
	\begin{equation}\label{FSVD2}
	\textbf{C}(s,u)=\sum_{i=1}^{\infty} \lambda_i \pmb{\psi}_i (s)\pmb{\psi}_i^\top (u).
	\end{equation}
This result is known as multivariate Mercer's Theorem, (see e.g. \cite{happ2016} Prop. 3). For any positive $i$, define an operator $\mathbfcal{X}_i:\mathbb{R}^K\rightarrow\mathbb{H}^L$, given by
	\begin{equation}\label{eq: elementary oprator}
	\mathbfcal{X}_i \pmb{a}:=\sum_{j=1}^K a_j (\pmb{\psi}_i\otimes \pmb{\psi}_i){\pmb x}_j= (\pmb{\psi}_i\otimes \pmb{\psi}_i)\sum_{j=1}^K a_j{\pmb x}_j.
	\end{equation}
We call $\mathbfcal{X}_i$ an \textit{elementary operator}. Note that $\mathbfcal{X}_i\in\mathbb{H}^{L\times K}$. Evaluating $\mathbfcal{X}_i \pmb{a}$ at a given point $s\in [0,1]$ is equivalent to the matrix product ${\bf X}_i(s)\pmb{a}$, where ${\bf X}_i(s)$ is an $L \times K$ matrix given by
	\begin{align}\label{eq:elementary mat}
	{\bf X}_i(s)&:=
	\begin{bmatrix} \langle\pmb{\psi}_i, {\pmb x}_1\rangle_{\mathbb{H}^L} \pmb{\psi}_i(s), \ldots, \langle\pmb{\psi}_i, {\pmb x}_K\rangle_{\mathbb{H}^L} \pmb{\psi}_i(s) \end{bmatrix} 
	\notag \\ &\ =
	\begin{bmatrix} (\pmb{\psi}_i\otimes \pmb{\psi}_i){\pmb x}_1(s), \ldots, (\pmb{\psi}_i\otimes \pmb{\psi}_i){\pmb x}_K(s) \end{bmatrix}.
	\end{align}
Note that, ${\bf X}_i(s)$'s can be considered as functional extension of the elementary matrices defined in \eqref{eq: primary}, where ${\bf X}_i(s)$ is projecting columns of ${\bf X}(s)$ into a spaced spanned by $\pmb{\psi_i}(s)$. 

\begin{proposition}\label{prop:rank1O}
The elementary operators $\mathbfcal{X}_i$'s are bounded operators of rank one. Furthermore $\mathbfcal{X}_i$'s decompose the trajectory operator $\mathbfcal{X}$ as 
	\begin{equation}\label{eq:elementary operators}
	\mathbfcal{X}=\sum_{i=1}^\infty \mathbfcal{X}_i.
	\end{equation}
\end{proposition}

The next theorem provides the SVD of the trajectory operator $\mathbfcal{X}$, to obtain the associated eigentriples ($\sqrt{\lambda_i},  \pmb{v}_i, \pmb{\psi}_i$) in the decomposition step.
\begin{theorem}
\label{thm:svd}
Let ${\{\pmb{\psi}_i\}_{i=1}^\infty}$ and ${\{\lambda_i\}_{i=1}^\infty}$ be the eigenelements of $\boldsymbol{\mathcal{S}}$ given in \eqref{eq:eigen}, and 
\begin{equation}
\boldsymbol{\mathcal{S}}^\dag:=\mathbfcal{X}^*\mathbfcal{X}=
\begin{bmatrix}
\sum_{i=1}^L\langle y_i, y_{i}\rangle & \cdots & \sum_{i=1}^L\langle y_i, y_{i+K-1}\rangle \\
\vdots & \ddots &\vdots \\
\sum_{i=1}^L\langle y_{i+K-1}, y_{i}\rangle & \cdots & \sum_{i=1}^L\langle y_{i+K-1}, y_{i+K-1}\rangle\\
\end{bmatrix}.
\end{equation}
Then, the SVD of the   trajectory operator $\mathbfcal{X}$ can be written as 
\begin{equation}
\label{eq: svd}
\mathbfcal{X} = \sum_{i=1}^\infty \sqrt{\lambda_i}\textbf{v}_i\otimes \pmb{\psi}_i  ,
\end{equation}
where 
$\textbf{v}_i=
\begin{pmatrix} \frac{\langle\pmb{\psi}_i, {\pmb x}_1\rangle_{\mathbb{H}^L}}{\sqrt{\lambda_i}} , \ldots,\frac{ \langle\pmb{\psi}_i, {\pmb x}_K\rangle_{\mathbb{H}^L}}{\sqrt{\lambda_i}} \end{pmatrix}^\top$. Furthermore for any $\pmb{a}\in\mathbb{R}^K$, using \eqref{eq: svd} we have 
\begin{equation}
\mathbfcal{X}\pmb{a} = \sum_{i=1}^\infty \sqrt{\lambda_i} \langle \textbf{v}_i, \pmb{a}\rangle_{\mathbb{R}^K}  \pmb{\psi}_i  ,
\end{equation}
where 
\begin{itemize}
\item[i)] $\{\lambda_i\}_{i=1}^\infty$ is the set of non-ascending eigenvalues of  $\boldsymbol{\mathcal{S}}^\dag,$ and
\item[ii)] $\pmb{v}_i$'s are the associated orthonormal eigenvectors of $\boldsymbol{\mathcal{S}}^\dag,$ satisfying $\mathbfcal{X}\pmb{v}_i = \sqrt{\lambda_i} \pmb{\psi}_i$.
\end{itemize}
\end{theorem}
We refer to $\sqrt{\lambda_i}$ as singular value, $\pmb{\psi}_i$ as left singular function, and $\pmb{v}_i$ as right singular vector associated to the $i^{th}$ component of the trajectory operator $\mathbfcal{X}$. The singular vectors, $\pmb{v}_i$'s, can be used to produce paired plots similar to the ones in the SSA literature \citep{golyandina2001analysis}.

\subsubsection*{Step 3. Grouping}
The grouping step is the procedure of rearranging and partitioning the elementary operators ${\mathbfcal X}_i$'s in \eqref{eq:elementary operators}. To do this, we mimic the approaches used in step 3 of Section \ref{ssa-method} for the univariate SSA and implement the equivalent functional version of those in \citet{rfssapackage}. Note that, in practice, we use a finite set of elementary operators, and one can consider a partition $\{{I_1}, {I_2}, \ldots, {I_m}\}$ of the set of indices such that we have the expansion
	\begin{equation}\label{eq:grouping}
	\mathbfcal{X}=\mathbfcal{X}_{I_1}+\mathbfcal{X}_{I_2}+\cdots+\mathbfcal{X}_{I_m}.
	\end{equation}

\subsubsection*{Step 4. Reconstruction}
At this step, for any given $q$ ($1\leq q\leq m$), we would like to use $\mathcal{T}^{-1}:\mathbb{H}_H^{L\times K} \rightarrow \mathbb{H}^N$ to transform back each operator $\mathbfcal{X}_{I_q}$ in \eqref{eq:grouping} to $\tilde{\bf y}_N^q$, and hence construct a functional version of the decomposition given in \eqref{ssa}. But since $\mathbfcal{X}_{I_q}\in\mathbb{H}^{L\times K}$, first it is necessary to project $\mathbfcal{X}_{I_q}$ to $\mathbb{H}_H^{L\times K}$. Note that $\mathbb{H}_H^{L\times K}$ is a closed subspace of $\mathbb{H}^{L\times K}$, therefore by Projection Theorem, there exist a unique $\mathbfcal{\tilde{X}}_{I_q}\in\mathbb{H}_H^{L\times K}$ such that
	$$\Vert \mathbfcal{X}_{I_q}-\mathbfcal{\tilde{X}}_{I_q} \Vert_\mathcal{F}^2 \leq \Vert \mathbfcal{X}_{I_q}-\tilde{\mathbfcal{Z}} \Vert_\mathcal{F}^2,\ \text{for any}\ \tilde{\mathbfcal{Z}}\in\mathbb{H}_H^{L\times K}.$$

To specify $\mathbfcal{\tilde{X}}_{I_q}$, we denote the elements of $\mathbfcal{X}_{I_q}$ and $\mathbfcal{\tilde{X}}_{I_q}$ by $[x_{i,j}^{q}]$ and $[\tilde{x}_{i,j}^{q}]$, respectively. Using Lemma \ref{lemma:lse}, it is easy to extend the diagonal averaging approach given by \citet{golyandina2001analysis} to $\mathbb{H}^{L\times K}$ and obtain $\tilde{x}_{i,j}^{q}$'s as following:
	\begin{equation}\label{fdiag-ave}
	\tilde{x}_{i,j}^{q}=\dfrac{1}{n_{s}}\sum_{(l,k): l+k =s}{x}_{l,k}^q,
	\end{equation}
where $s=i+j$ and $n_s$ stands for the number of $(l,k)$ pairs such that $l+k=s$. Denote this projection by $\Pi_\mathbb{H}:\mathbb{H}^{L\times K}\rightarrow \mathbb{H}^{L\times K}_H$, and set $\mathbfcal{\tilde{X}}_{I_q}=\Pi_\mathbb{H} \mathbfcal{X}_{I_q}$. Now we can define $\tilde{\bf y}_N^q=\mathcal{T}^{-1}\tilde{\mathbfcal{X}}_{I_q}$, and reconstruct the functional time series.

\subsection{Separability}\label{subsec: Separability}
Let $\textbf{y}_N=\textbf{y}_N^{(1)}+\textbf{y}_N^{(2)}$, where $\textbf{y}_N^{(i)}=\lbrace y_1^{(i)},\ldots, y_N^{(i)}\rbrace$, $i=1,2$, are FTS. Using a fixed window length $L$, for each series $\textbf{y}_N^{(i)}$, denote $\lbrace{\pmb x}_{k}^{(i)}\rbrace_{k=1}^K$ as a sequence of functional lagged vectors, and $\mathcal{L}^{(i)}$ as the linear space spanned by $\lbrace{\pmb x}_{k}^{(i)}\rbrace_{k=1}^K$. Analogous to univariate SSA, separability of the series $\textbf{y}_N^{(1)}$ and $\textbf{y}_N^{(2)}$ is equivalent to $\mathcal{L}^{(1)} \bot \ \mathcal{L}^{(2)}$, which is same as $\langle {\pmb x}_{k}^{(1)} , {\pmb x}_{k^\prime}^{(2)}\rangle_{\mathbb{H}^L}=0$, for all $k, k^\prime = 1,\ldots, K$. Furthermore, a necessary condition for separability can be defined based on w-correlation measure. To do this, consider the weighted inner product of two series $\textbf{y}_N^{(1)}$ and $\textbf{y}_N^{(2)}$ as 
	\begin{equation}\label{winn}
	\left\langle \textbf{y}_N^{(1)} , \textbf{y}_N^{(2)} \right\rangle_w=\sum_{i=1}^N w_i \langle y_i^{(1)} , y_i^{(2)}\rangle,
	\end{equation}
where $w_i=\min\{i,L,N-i+1\}$. We call the series $\textbf{y}_N^{(1)}$ and $\textbf{y}_N^{(2)}$ w-orthogonal if 
	\begin{equation}
	\left\langle \textbf{y}_N^{(1)} , \textbf{y}_N^{(2)} \right\rangle_w=0.
	\end{equation}

\begin{theorem}\label{thm:fSep}
If the series $\textbf{y}_N^{(1)}$ and $\textbf{y}_N^{(2)}$ are separable, then they are w-orthogonal. 
\end{theorem}

Also, to quantify the degrees of separability of two FTS, the functional version of the w-correlation measure can be obtained by replacing the new definition of the weighted inner product \eqref{winn} into \eqref{wcor}.

\section{Implementation Strategy}\label{sec:Practical Imp.}
In practice, functional data are being recorded discretely and then converted to functional objects using proper smoothing techniques. We refer to \cite{ramsay2007} for more details on preprocessing the raw data. Let $\left\lbrace \nu_i \right\rbrace_{ i\in \mathbb{N}}$ be a known basis system (not necessarily orthogonal) of the space $\mathbb{H}$. Each functional observation in $\mathbb{H}$ can be projected into subspace ${\mathbb{H}}_d:=sp \left\lbrace \nu_i \right\rbrace$, where $d$ can be determined by variety of techniques (e.g. cross-validation). Therefore, each $y_j\in \mathbb{H}_d$ is uniquely represented by
	\begin{equation}\label{yj}
	y_j(s) = \sum_{i=1}^d a_{i,j}\nu_i(s),\ j=1, \ldots, N,\ s \in [0,1].
	\end{equation}

Let us define quotient sequence, $q_k$, and reminder sequence, $r_k$, by
	\begin{equation}\label{k}
	k=(q_k-1)L+r_k,\ 1\leq r_k\leq L,\ 1\leq q_k\leq d.
	\end{equation}
Note that for any given $k$ ($1\leq k \leq Ld$), one may use \eqref{k} to determine $q_k$ and $r_k$ uniquely, so these sequences are well defined. Now, consider the objects ${\pmb \phi}_{k} \in \mathbb{H}_d^L$, as a vector of length $L$ with all coordinates are zero except $r_k$-th, which is $\nu_{q_k}$.

\begin{lemma}\label{basis-lemma}
The sequence $\{{\pmb \phi}_{k}\}_{k=1}^{Ld}$ is a basis system for $\mathbb{H}_d^L$, where $\mathbb{H}_d^L$ is the Cartesian product of $L$ copies of $\mathbb{H}_d$.
\end{lemma}

Using Lemma \ref{basis-lemma}, each element ${\pmb z}\in \mathbb{H}_d^L$ admits a unique representation 
	\begin{equation}\label{A}
	{\pmb z} = \sum_{i=1}^{Ld} \langle {\pmb z}, \widetilde{{\pmb\phi}}_{i} \rangle_{\mathbb{H}^L}{\pmb \phi}_{i}:={\mathbfcal F} {{\mathbf c_{\pmb z}}},
	\end{equation}
where ${\mathbf c_{\pmb z}}=\left(\langle {\pmb z}, \widetilde{{\pmb\phi}}_{1} \rangle_{\mathbb{H}^L},\ldots ,\langle {\pmb z}, \widetilde{{\pmb\phi}}_{Ld} \rangle_{\mathbb{H}^L}\right)^\top$ corresponds to ${\pmb z}$,\ ${\mathbfcal{F}}:\mathbb{R}^{Ld}\rightarrow \mathbb{H}_d^L$ belongs to $\mathbb{H}^{Ld\times L}_d$, and $ \widetilde{{\pmb\phi}}_{i}$ is the dual basis of ${\pmb \phi}_{i}$. Note that, in the special case, when $\nu_i$'s are orthonormal (so ${\pmb \phi}_{i}$'s are), $\widetilde{{\pmb\phi}}_{i} = {\pmb \phi}_{i}$ \citep[see][for more details]{christensen1995}. Applying the linear operator $\mathbfcal{S}$, defined in \eqref{eq: s-operator}, on \eqref{A} implies
	\begin{align}
	{\mathbfcal{S}}\pmb{z}&=\sum_{i=1}^{Ld} \langle {\pmb z}, \widetilde{{\pmb\phi}}_{i} \rangle_{\mathbb{H}^L}{\mathbfcal{S}}{\pmb \phi}_{i}
	=\sum_{i=1}^{Ld} \sum_{j=1}^{Ld} \langle {\pmb z}, \widetilde{{\pmb\phi}}_{i} \rangle_{\mathbb{H}^L}\langle {\mathbfcal{S}}{\pmb \phi}_{i},
	 \widetilde{{\pmb\phi}}_{j} \rangle_{\mathbb{H}^L} {\pmb \phi}_{j}={\mathbfcal F}{\mathbf{S}}{\mathbf c_{\pmb z}},
	\end{align}
where ${\mathbf{S}}^\top=\left[\langle {\mathbfcal{S}}{\pmb \phi}_{i}, \widetilde{{\pmb\phi}}_{j} \rangle_{\mathbb{H}^L}\right]_{i,j=1}^{Ld}$. We call ${\mathbf{S}}$ the corresponding matrix of ${\mathbfcal{S}}$. 

\begin{lemma}\label{lemma1}
Let ${\pmb z}$ be a functional object in $\mathbb{H}_d^L$, then ${\mathbfcal{S}}\pmb{z}=\lambda{\pmb z}$ if and only if ${\mathbf{S}}{\mathbf c_{\pmb z}}=\lambda {\mathbf c_{\pmb z}}$.
\end{lemma}

\begin{theorem}\label{s-elementThm}
Suppose the Gram matrix $\mathbf{G}:=\left[ \langle {\pmb \phi}_{i}, {\pmb \phi}_{j} \rangle_{\mathbb{H}^L} \right]_{i,j=1}^{Ld}$. Then the following holds:
	\begin{itemize}
	\item[(i)] $\mathbf{G}=\left[ \delta_{r_i,r_j}\langle \nu_{q_i} , \nu_{q_j} \rangle \right]_{i,j=1}^{Ld}$.
	\item[(ii)] $\left\langle \mathbfcal{S}{\pmb \phi}_{i} , {\pmb \phi}_{j}\right\rangle_{\mathbb{H}^L}=\sum_{m=1}^K\langle y_{r_i+m-1} , \nu_{q_i}\rangle\langle y_{r_j+m-1} , \nu_{q_j}\rangle$.
	\item[(iii)] ${\mathbf{S}}_0:=\left[\langle {\mathbfcal{S}}{\pmb \phi}_{i}, {\pmb \phi}_{j} \rangle_{\mathbb{H}^L}\right]_{i,j=1}^{Ld}$ is a symmetric matrix.
	\item[(iv)] ${\mathbf{S}}=\mathbf{G}^{-1}\mathbf{S}_0$. 
	\end{itemize}
\end{theorem}

Now we have the recipes to proceed with the following algorithm and obtain the eigenfunctions of $\mathbfcal{S}$, $\pmb \psi_i$'s, used in the decomposition step. For a given set of basis $\{\nu_i\}_{i=1}^d$, and a FTS, $\textbf{y}_N$:
	\begin{itemize}
	\item Use Theorem \ref{s-elementThm} to compute the matrices $\mathbf{G},\ \mathbf{S}_0$, and $\mathbf{S}$.
	\item Use the eigendecomposition of $\mathbf{S}$ to obtain eigenpairs $(\lambda_i, {\mathbf c_{{\pmb{\psi}}_i}})$ for $i=1,\ldots, Ld$.
	\item Use \eqref{A} and Lemma \ref{lemma1} to obtain $\pmb{\psi_i}$'s, eigenfunctions of $\mathbfcal{S}$.
	\end{itemize}
Now, one can use ${{\pmb{\psi}}_i}$'s to decompose the FTS to elementary operators $\mathbfcal{X}_i$'s. Note that, in practice, we represent the elementary operators in matrix form. Therefore, one may observe that the equivalent functional elementary matrix ${\bf X}_i(s)$, given in \eqref{eq:elementary mat}, is just a projection of the functional lagged vectors $\{{\pmb x}_j(s)\}_{j=1}^K$, given in \eqref{flvec}, onto ${{\pmb{\psi}}_i}$. Furthermore, in the diagonal averaging step, we can incorporate the averaging over the associate basis coefficients of $x_{l,k}^q$ in \eqref{fdiag-ave} to obtain the respective basis coefficients for $\tilde{x}_{i,j}^q$. For more details see \citet{rfssapackage}.

\section{Numerical study}\label{simul-real}
In this section, first, we present a simulation study to elaborate the use of the FSSA compared with dFPCA and MSSA under different scenarios. To do so, we utilize the implementation of the proposed model that is available as an \proglang{R} package named \pkg{Rfssa} in the {CRAN} repository \citep{rfssapackage}. We also use the \pkg{freqdom.fda} \citep{hormann2015} and \pkg{Rssa} \citep{rssapackage} packages to obtain the dFPCA and MSSA results. In the second subsection, we analyze a remote sensing data using \pkg{Rfssa} and provide some visualization tools that come handy in the grouping step.
 
We developed a shiny app, included in the \pkg{Rfssa} package, also available at \url{https://fssa.shinyapps.io/fssa/}, to demonstrate and reproduce different aspects of the simulation setup. Furthermore, it can be used to compare the results of dFPCA, MSSA, and FSSA on the remote sensing data, call center dataset or any other FTS, provided by the end-user.

\subsection{Simulation study}\label{simul}
For the simulation setup, consider the functional time series of lengths $N=50,100, 150$ and $200$ which are observed in $n=100$ fixed equidistant discrete points on $[0,1]$ from the following model: 
	\begin{equation}\label{eq:mainmodel}
	Y_t\left(s_i\right)=m_t(s_i)+X_t\left(s_i\right),\ s_i \in [0,1], i=1,\ldots,n, \text{and } t=1, \dots, N. \
	\end{equation}
A cubic B-spline basis functions with 15 degrees of freedom is used to convert $\{Y_t(s_i)\}$'s into smooth (continuous) functional curves. In this model, $m_t(s)$ is considered to be a periodic component defined as 
	\begin{equation}\label{eq:Trend}
	m_t(s)=e^{s^2} \cos\left(2\pi \omega t\right)+\cos(4\pi s) \sin\left(2\pi \omega t\right),
	\end{equation}
where $\omega$ is the model frequency with three different values ($\omega=0, 0.1$ and $0.25$).

The $X_t(s)$ in \eqref{eq:mainmodel} is a stochastic term that is generated under four different settings with an increasing trend in complexity. In the first setting, we consider $\{X_t(s_i),t=1,\ldots, N\ \text{and}\ i=1,\ldots,n \}$ are drawn from an independent Gaussian White Noise (GWN) process with zero mean and standard deviation equal to $0.1$. It is expected to obtain an acceptable performance from FPCA for reconstructing the FTS in the first setting as intuitively FPCA outperforms under this ideal framework \citep[see][for more details]{maadooliat2015integrating}.

In the remaining three settings, the $\{X_t(s)\}$ processes are simulated from a functional autoregressive model of order 1, $FAR(1)$, defined by 
	\begin{equation}\label{eq:FAR(1)}
	X_t(s)=\Psi X_{t-1}(s)+\varepsilon_t(s),
	\end{equation}
where $\Psi$ is an integral operator with a parabolic kernel as follow
	\begin{equation*}\label{eq:parabolic}
	\psi(s,u)=\gamma_0\left(2-(2s-1)^2-(2u-1)^2\right).
	\end{equation*}
The constant $\gamma_0$ is chosen such that the Hilbert-Schmidt norm defined by
	$$\Vert \Psi \Vert_\mathcal{S}^2=\int_0^1\int_0^1|\psi(s,u)|^2ds du,$$
acquires the values $\Vert \Psi \Vert_\mathcal{S}^2=0, 0.5$, and $0.9$, for the remaining three settings, respectively. In these settings, the white noise terms $\varepsilon_t(s)$ are considered as independent trajectories of the standard Brownian motion over the interval $[0,1]$. It is worth to note that as we increase the Hilbert-Schmidt norm, $\Vert \Psi \Vert_\mathcal{S}^2$, in the FAR(1) models, the dependency structure of consecutive FTS gets more twisted, and we expect it would be more challenging to reconstruct the true structures, $\{X_t(s)\}$. 
 
To compare the performance of FSSA and MSSA, we further consider three window length parameters ($L=20, 30$ and $40$) in our simulation setup. For the sake of consistency in all of the reconstruction procedures (dFPCA, MSSA and FSSA), we use the first two leading eigen-components. As a measure of goodness of fit, we use the Root Mean Square Error (RMSE) defined as: 
	$$RMSE= \sqrt {\frac{1}{N\times n}\sum\limits_{t=1}^N \sum_{i=1}^n \left(Y_t(s_i)-\hat{Y}_t(s_i)\right)^2},$$
where $\hat{Y}_t(s_i)$ is the FTS reconstructed by each method. We repeat each setting $1000$ times and report the mean of the RMSE's in Table \ref{tab:sim}. 

By comparing the results in Table \ref{tab:sim}, it can be seen that FSSA outperform dFPCA in different scenarios. This may not be surprising, as the main task of dFPCA is dimension reduction. Except for the first setting, MSSA also outperforms dFPCA significantly. Furthermore, FSSA performs better than MSSA in most of the cases except the case where the length of the FTS is small ($N=50$) and the window size, $L$, is getting closer to $N$. However, it is clear that FSSA is the optimal method for reconstructing the longer FTS ($N\geq100$). 

For all methods, RMSE decreases as the length of the series increases. For two smaller frequencies ($\omega=0\ \textrm{and}\ 0.1$), the average of RMSE increases as the noise structure becomes more complex in settings 1 through 4 while it decreases for $\omega=0.25$. This might be happening due to the unpredicted cross-correlation of the functional noise structures and the periodic form of FTS. 

\begin{figure}[!h]
\begin{center}
\includegraphics[scale=.8]{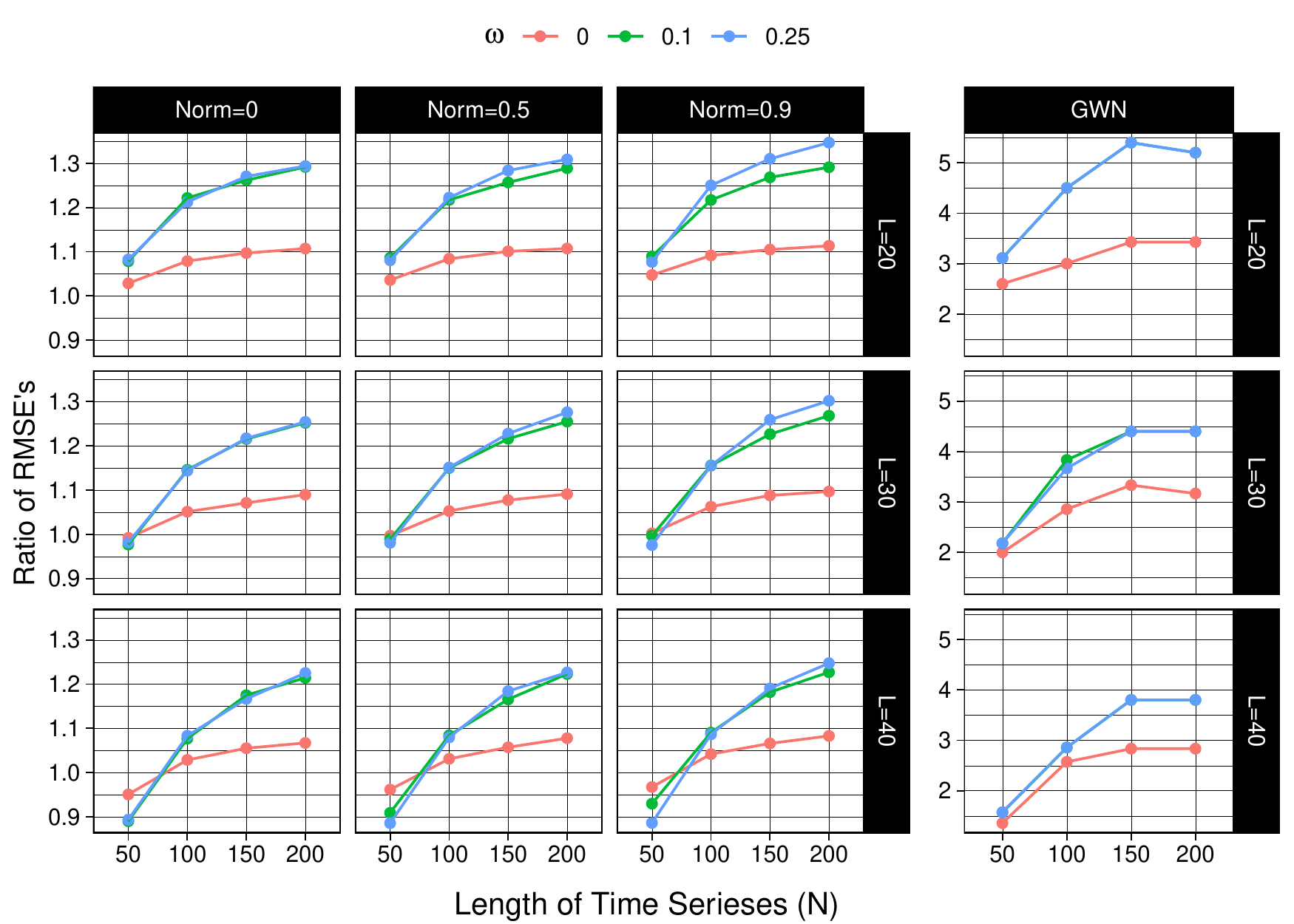}
\caption{Ratio of RMSE of MSSA to FSSA in simulation study with 1000 repetition.}
\label{fig:ratioRMSE}
\end{center}
\end{figure}

The efficiency of MSSA and FSSA for different window lengths ($L$), time series lengths ($N$) and frequencies ($\omega$), the ratio of RMSE of MSSA to FSSA is examined in Figure \ref{fig:ratioRMSE}. The overall pattern confirms the improvement in RMSE for FSSA as the time series get longer (larger $N$). Overall, as $L$ is increasing, the pattern of ratio of RMSE's remains unchanged. Although as the window length becomes larger, either the improvement diminishes for longer FTS, or disappears (or reverses) for smaller $N$. It is also worth to note that in setting 1 (GWN), based on the right panel of Figure \ref{fig:ratioRMSE} and Table \ref{tab:sim}, the FSSA dominates the other two methods in all combinations of parameters with a better efficiency scale.

\begin{table}[p]
\centering
\def\arraystretch{0.57}
\begin{scriptsize}
\resizebox{\columnwidth}{!}{
\begin{tabular}{lcccccccccc}
 \hline
Model & $\omega$ & N & dFPCA & & MSSA & && & FSSA & \\ 
 \cline{5-7} \cline{9-11}&&&&$L=20$&$L=30$&$L=40$&&$L=20$&$L=30$&$L=40$\\
\hline
\multirow{12}{*}{\shortstack{Setting 1\\ GWN}} & 0.00 & 50 & 0.092 & 0.026 & 0.022 & 0.019 && 0.010 & 0.011 & 0.014 \\ 
&& 100 & 0.088 & 0.024 & 0.020 & 0.018 && 0.008 & 0.007 & 0.007 \\ 
&& 150 & 0.074 & 0.024 & 0.020 & 0.017 && 0.007 & 0.006 & 0.006 \\ 
&& 200 & 0.071 & 0.024 & 0.019 & 0.017 && 0.007 & 0.006 & 0.006 \\ \cline{2-11} 
& 0.10 & 50 & 0.062 & 0.028 & 0.024 & 0.022 && 0.009 & 0.011 & 0.014 \\ 
&& 100 & 0.045 & 0.027 & 0.023 & 0.020 && 0.006 & 0.006 & 0.007 \\ 
&& 150 & 0.037 & 0.027 & 0.022 & 0.019 && 0.005 & 0.005 & 0.005 \\ 
&& 200 & 0.033 & 0.026 & 0.022 & 0.019 && 0.005 & 0.005 & 0.005 \\ \cline{2-11} 
& 0.25 & 50 & 0.046 & 0.028 & 0.024 & 0.022 && 0.009 & 0.011 & 0.014 \\ 
&& 100 & 0.038 & 0.027 & 0.022 & 0.020 && 0.006 & 0.006 & 0.007 \\ 
&& 150 & 0.032 & 0.027 & 0.022 & 0.019 && 0.005 & 0.005 & 0.005 \\ 
&& 200 & 0.030 & 0.026 & 0.022 & 0.019 && 0.005 & 0.005 & 0.005 \\ \cline{1-11} 
\multirow{12}{*}{\shortstack{Setting 2 \\ $\Vert \Psi \Vert_\mathcal{S}^2=0$ }}& 0.00 & 50 & 1.008 & 0.251 & 0.246 & 0.270 && 0.244 & 0.248 & 0.284 \\ 
&& 100 & 0.993 & 0.205 & 0.185 & 0.179 && 0.190 & 0.176 & 0.174 \\ 
&& 150 & 0.974 & 0.192 & 0.166 & 0.153 && 0.175 & 0.155 & 0.145 \\ 
&& 200 & 0.960 & 0.186 & 0.158 & 0.143 && 0.168 & 0.145 & 0.134 \\ \cline{2-11} 
& 0.10 & 50 & 0.636 & 0.220 & 0.211 & 0.235 && 0.204 & 0.216 & 0.264 \\ 
&& 100 & 0.633 & 0.193 & 0.165 & 0.154 && 0.158 & 0.144 & 0.143 \\ 
&& 150 & 0.633 & 0.188 & 0.158 & 0.141 && 0.149 & 0.130 & 0.120 \\ 
&& 200 & 0.632 & 0.186 & 0.154 & 0.136 && 0.144 & 0.123 & 0.112 \\ \cline{2-11} 
& 0.25 & 50 & 0.636 & 0.223 & 0.214 & 0.235 && 0.206 & 0.218 & 0.263 \\ 
&& 100 & 0.633 & 0.194 & 0.167 & 0.156 && 0.160 & 0.146 & 0.144 \\ 
&& 150 & 0.633 & 0.188 & 0.157 & 0.140 && 0.148 & 0.129 & 0.120 \\ 
&& 200 & 0.632 & 0.185 & 0.153 & 0.136 && 0.143 & 0.122 & 0.111 \\ \cline{1-11} 
\multirow{12}{*}{\shortstack{Setting 3\\ $\Vert \Psi \Vert_\mathcal{S}^2=0.5$}} & 0.00 & 50 & 0.841 & 0.316 & 0.310 & 0.350 && 0.305 & 0.311 & 0.364 \\ 
&& 100 & 0.818 & 0.271 & 0.240 & 0.231 && 0.250 & 0.228 & 0.224 \\ 
&& 150 & 0.806 & 0.262 & 0.223 & 0.203 && 0.238 & 0.207 & 0.192 \\ 
&& 200 & 0.801 & 0.258 & 0.216 & 0.194 && 0.233 & 0.198 & 0.180 \\ \cline{2-11} 
& 0.10 & 50 & 0.696 & 0.277 & 0.267 & 0.300 && 0.255 & 0.270 & 0.330 \\ 
&& 100 & 0.709 & 0.241 & 0.207 & 0.193 && 0.198 & 0.180 & 0.178 \\ 
&& 150 & 0.710 & 0.235 & 0.197 & 0.176 && 0.187 & 0.162 & 0.151 \\ 
&& 200 & 0.710 & 0.232 & 0.192 & 0.170 && 0.180 & 0.153 & 0.139 \\ \cline{2-11} 
& 0.25 & 50 & 0.707 & 0.214 & 0.205 & 0.225 && 0.198 & 0.209 & 0.254 \\ 
&& 100& 0.707 & 0.187 & 0.160 & 0.149 && 0.153 & 0.139 & 0.138 \\ 
&& 150 & 0.705 & 0.181 & 0.151 & 0.135 && 0.141 & 0.123 & 0.114 \\ 
&& 200 & 0.705 & 0.178 & 0.148 & 0.130 && 0.136 & 0.116 & 0.106 \\ \cline{1-11} 
\multirow{12}{*}{\shortstack{Setting 4\\ $\Vert \Psi \Vert_\mathcal{S}^2=0.9$ }}& 0.00 & 50 & 0.926 & 0.486 & 0.477 & 0.535 && 0.464 & 0.476 & 0.553 \\ 
&& 100 & 0.915 & 0.441 & 0.391 & 0.373 && 0.404 & 0.368 & 0.358 \\ 
&& 150 & 0.901 & 0.432 & 0.371 & 0.339 && 0.391 & 0.341 & 0.318 \\ 
&& 200 & 0.823 & 0.432 & 0.363 & 0.326 && 0.388 & 0.331 & 0.301 \\ \cline{2-11} 
& 0.10 & 50 & 0.830 & 0.341 & 0.330 & 0.384 && 0.313 & 0.331 & 0.413 \\ 
&& 100 & 0.831 & 0.286 & 0.245 & 0.229 && 0.235 & 0.212 & 0.210 \\ 
&& 150 & 0.832 & 0.279 & 0.233 & 0.208 && 0.220 & 0.190 & 0.176 \\ 
&& 200 & 0.829 & 0.275 & 0.227 & 0.200 && 0.213 & 0.179 & 0.163 \\ \cline{2-11} 
& 0.25 & 50 & 0.829 & 0.210 & 0.200 & 0.226 && 0.195 & 0.205 & 0.255 \\ 
&& 100 & 0.829 & 0.175 & 0.149 & 0.138 && 0.140 & 0.129 & 0.127 \\ 
&& 150 & 0.829 & 0.169 & 0.141 & 0.125 && 0.129 & 0.112 & 0.105 \\ 
&& 200 & 0.833 & 0.167 & 0.138 & 0.121 && 0.124 & 0.106 & 0.097 \\ 
\hline
\end{tabular}
}
\end{scriptsize}

\caption{The mean of RMSE for 1000 generation of the simulated model by dFPCA, MSSA and FSSA approaches. \label{tab:sim}}
\end{table}

\subsection{Application to Remote Sensing Data}\label{sec:NDVI}

Tracking changes in vegetation over time has become of interest to researchers that want to preserve wildlife. One technique used to track how much vegetation is present in a region is through field surveys. The problem with this technique is that it is difficult to implement especially in low population areas \citep{Panuju2012}. The use of remote sensing data in this context is preferred in order to detect man-made or natural changes in vegetation. One source of remote sensing data is from NASA's MODerate-resolution Imaging Spectroradiometer (MODIS) satellite which provides images, twice daily, of regions around the globe at varying spatial resolutions \citep{Tuck2014}. Normalized Difference Vegetation Index (NDVI) is a commonly used pixel-wise index in MODIS satellite images. The NDVI values are bounded between zero and one, where index values that are closer to one indicate that more vegetation is present and smaller values indicate the absence of vegetation \citep{Tuck2014}. 

Many studies have used the spectral NDVI measure and its variants in order to remotely track changes in vegetation over time \citep{Lambin1999}. The temporal average, and temporal variability of NDVI images have been used as explanatory variables for the number of different types of vegetation present in many regions \citep{Tuck2014}. Also, \cite{Panuju2012} used the maximum value of NDVI images, taken of the Jambi province in Indonesia, to form a time series. The resulted time series was then analyzed using an X12-ARIMA model in order to identify trend and seasonal changes in woody vegetation \citep{Panuju2012}. While these statistics (e.g. maximum and average) have seen some success in tracking the changes of the NDVI images, it may fail capturing the distribution of the vegetation. Therefore, one may seek for a more comprehensive measure (e.g. a FTS) to describe the distribution of the NDVI images.

\begin{figure}[!h]
\begin{center}
	\begin{subfigure}[b]{0.4\textwidth}
		\includegraphics[page=1,width=\textwidth]{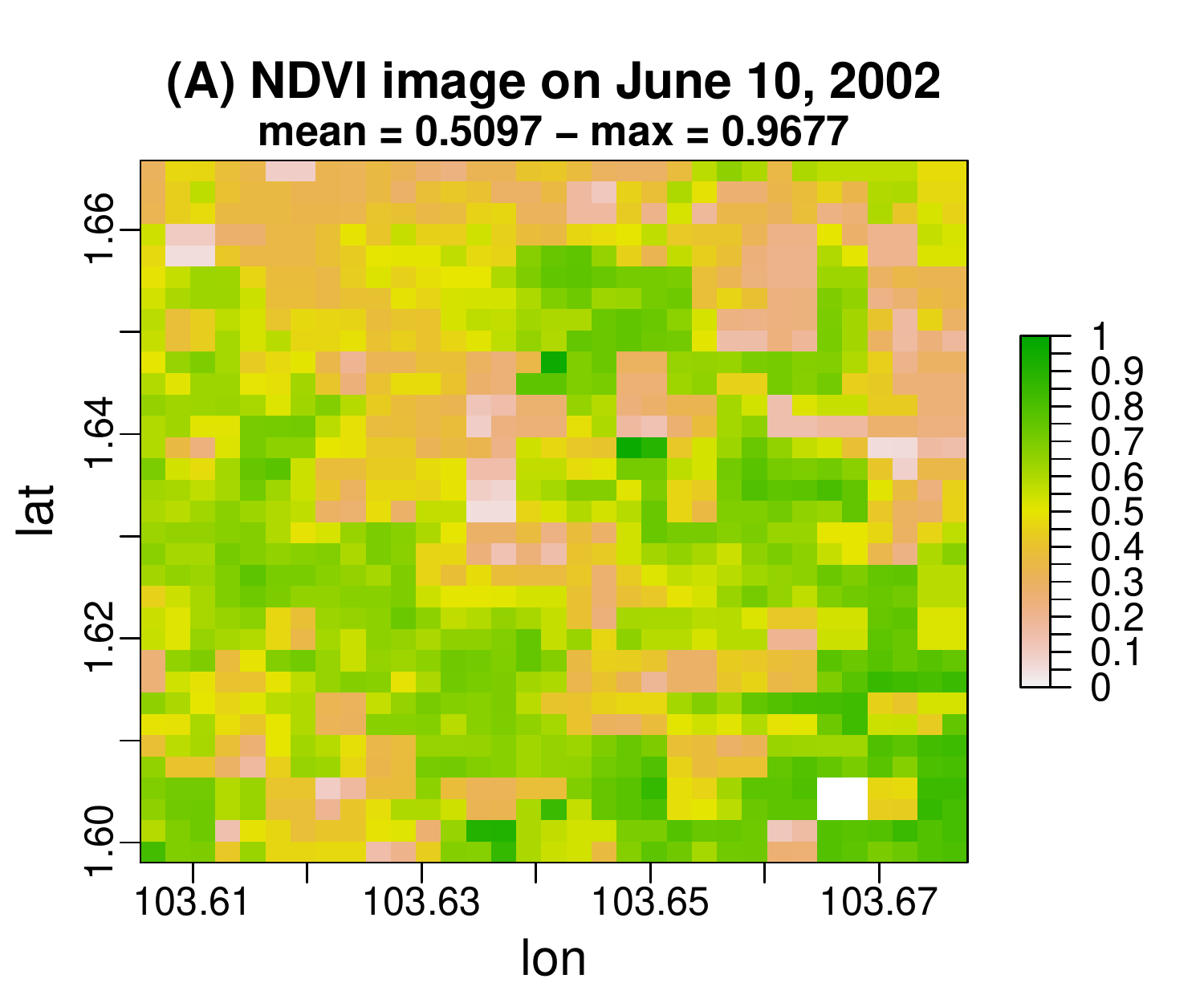}
	\end{subfigure}	
	\begin{subfigure}[b]{0.4\textwidth}
		\includegraphics[page=2,width=\textwidth]{NDVI-justification}
	\end{subfigure}	
	\begin{subfigure}[b]{0.4\textwidth}
		\includegraphics[page=3,width=\textwidth]{NDVI-justification}
	\end{subfigure}	
	\begin{subfigure}[b]{0.4\textwidth}
		\includegraphics[page=4,width=\textwidth]{NDVI-justification}
	\end{subfigure}	
 \caption{NDVI images and the associated KDEs on June 10 2002, and  June 10 2019 in the Jambi region.}
 \label{fig:ndvi_densities}
\end{center}
\end{figure}

Here we use $448$ NDVI images taken in $16$ days increments between February $18$, $2002$ to July $28$, $2019$. The images have a spatial resolution of $250$ meter and are of a square region just outside of the city of Jambi, Indonesia between $103.61^{\circ}E - 103.68^{\circ}E$ and $1.60^{\circ}S - 1.67^{\circ}S$. Figure \ref{fig:ndvi_densities} shows the respective NDVI images taken on June 10, 2002 and June 10, 2019. It is interesting to note that, although the respective NDVI images are not similar, the sample means of the NDVI values are very close (differ by only about 0.0032 unit). As it is shown in Figure \ref{fig:ndvi_densities}, the kernel density estimates (KDEs) are much more informative in representing the distribution of the NDVI images.

\begin{figure}[!h]
	\begin{subfigure}[b]{0.32\textwidth}
		\includegraphics[width=\textwidth]{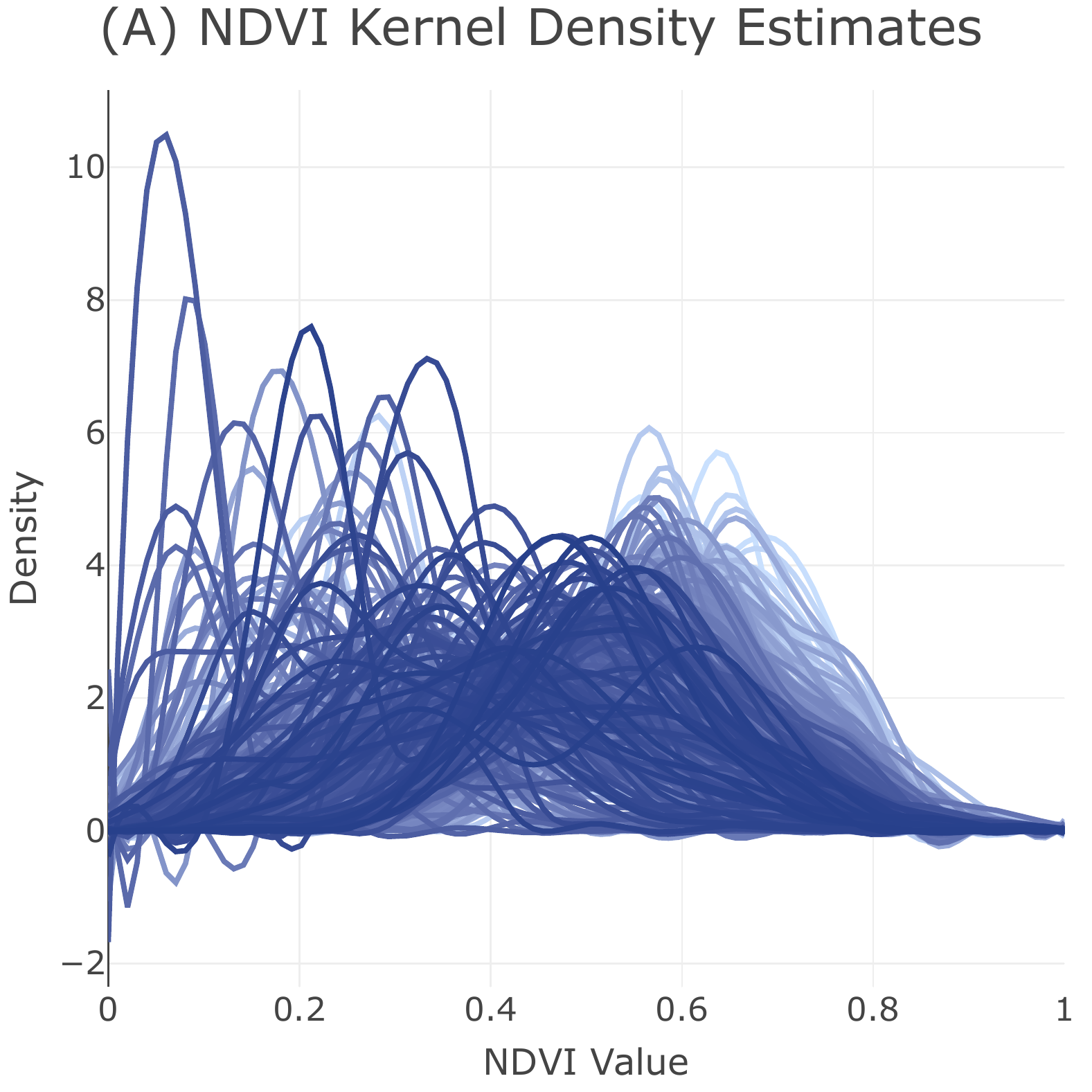}
		\label{fig:NDVI-A}
	\end{subfigure}	
	\begin{subfigure}[b]{0.33\textwidth}
		\includegraphics[page=1,width=\textwidth]{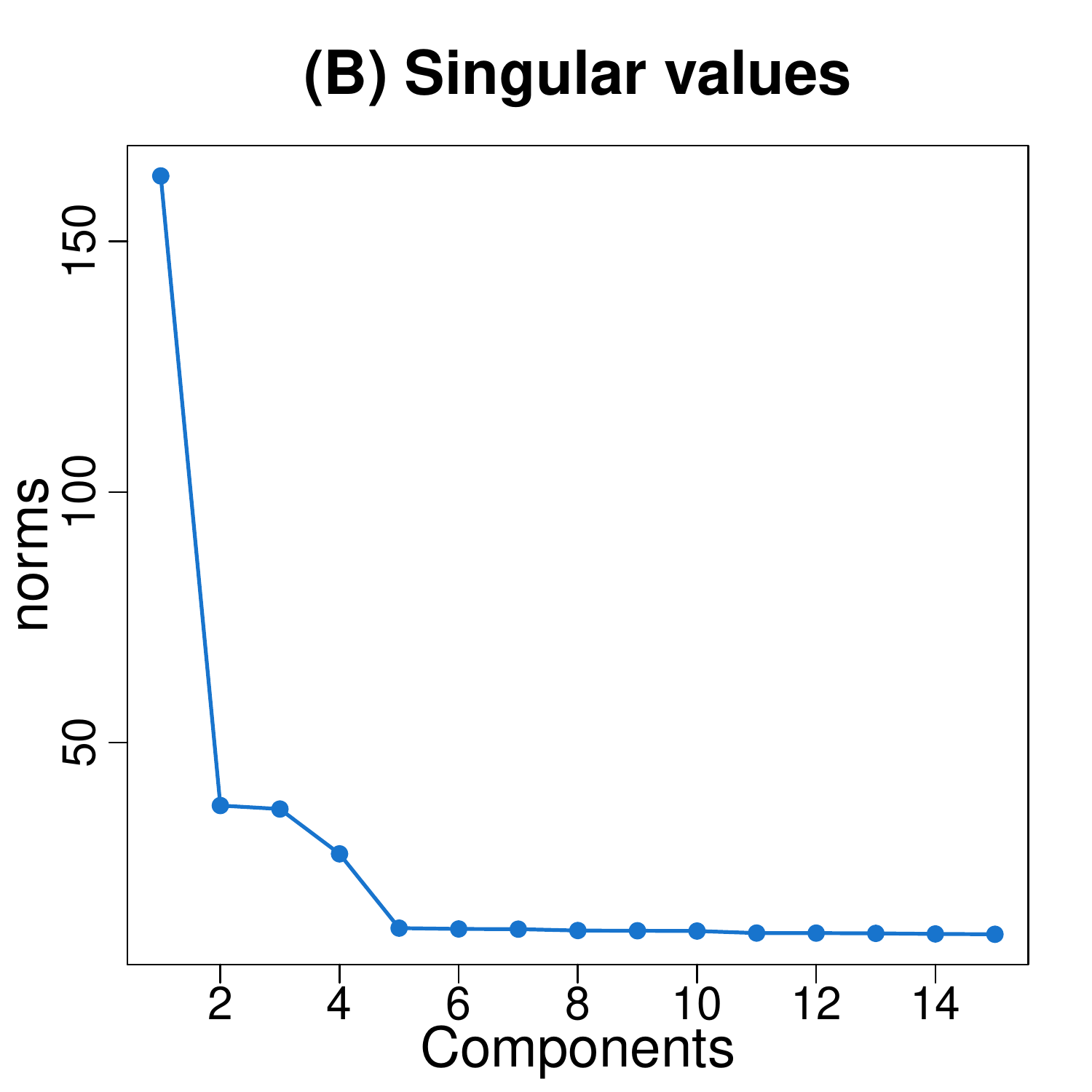}
		\label{fig:NDVI-B}
	\end{subfigure}
	\begin{subfigure}[b]{0.33\textwidth}
		\includegraphics[page=2,width=\textwidth]{NDVI-rest}
		\label{fig:NDVI-C}
	\end{subfigure}
	\begin{subfigure}[b]{0.33\textwidth}
		\includegraphics[page=3,width=\textwidth]{NDVI-rest}
		\label{fig:NDVI-D}
	\end{subfigure}
	\begin{subfigure}[b]{0.33\textwidth}
		\includegraphics[page=4,width=\textwidth]{NDVI-rest}
		\label{fig:NDVI-E}
	\end{subfigure}
	\begin{subfigure}[b]{0.32\textwidth}
		\includegraphics[page=5,width=\textwidth]{NDVI-rest}
		\label{fig:NDVI-F}
	\end{subfigure}
\caption{The KDEs of the $448$ NDVI images, plus the FSSA plots for the grouping steps of the NDVI dataset.}
\label{fig:ndvi_group}
\end{figure}

We follow the Silverman's rule of thumb \citep{Silverman1986} for bandwidth selection and obtain the KDEs. Then we project the results onto a functional space spanned by a cubic B-spline basis, selected via the GCV criterion. Figure \ref{fig:ndvi_group}(A) shows the projected KDEs onto the B-spline space. We pass the results ($448$ FTS) as input to the FSSA algorithm with the lag, $L=45$, and study the behavior of the NDVI images over almost two decade, using the \textbf{Rfssa} package.

According to the subplots in Figures \ref{fig:ndvi_group}(B-C: the singular values and w-correlation plots), a suitable partition would be grouping the first and fourth components separately, plus the second and third components jointly ($G=\{1, 2$--$3,4\}$). The remaining subplots in Figures \ref{fig:ndvi_group}(D-F: the right singular vectors and left singular functions) indicate the first component captures mean behavior, the second and third capture annual behavior, and the fourth captures trend.

\begin{figure}[!h]
\begin{center}
	\begin{subfigure}[b]{0.4\textwidth}
		\includegraphics[width=\textwidth]{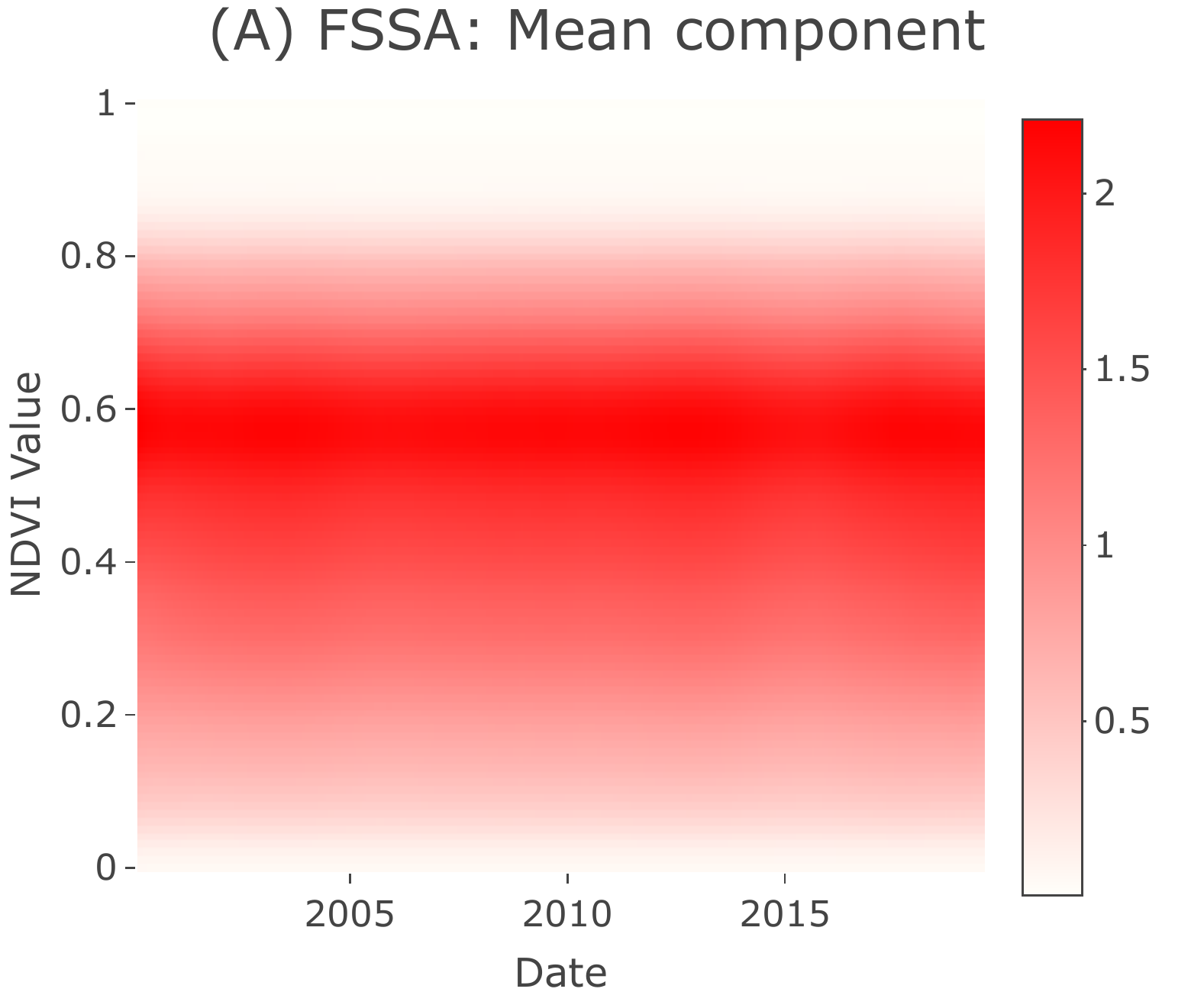}
	\end{subfigure}	
	\begin{subfigure}[b]{0.4\textwidth}
		\includegraphics[width=\textwidth]{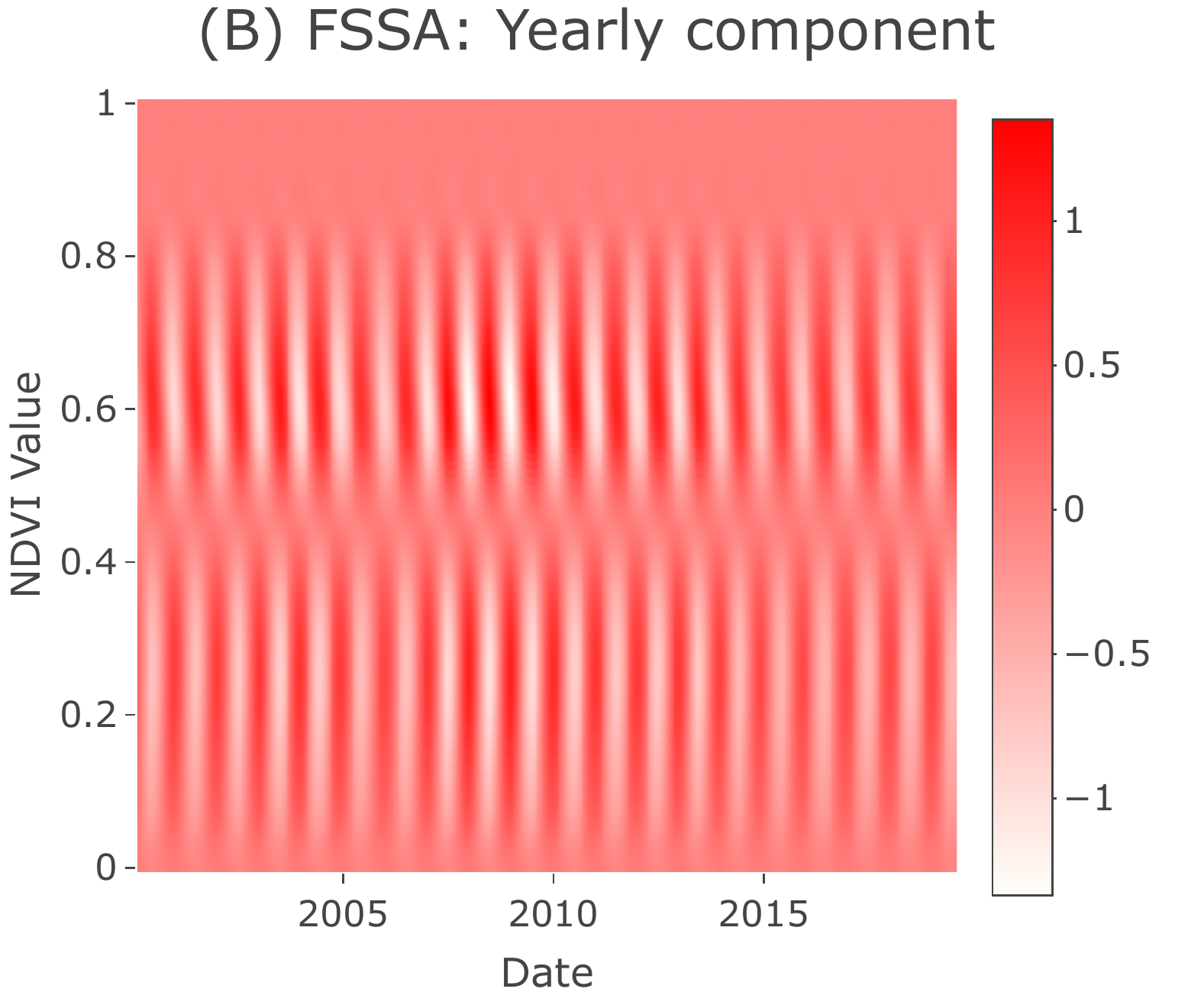}
	\end{subfigure}	
	\begin{subfigure}[b]{0.4\textwidth}
		\includegraphics[width=\textwidth]{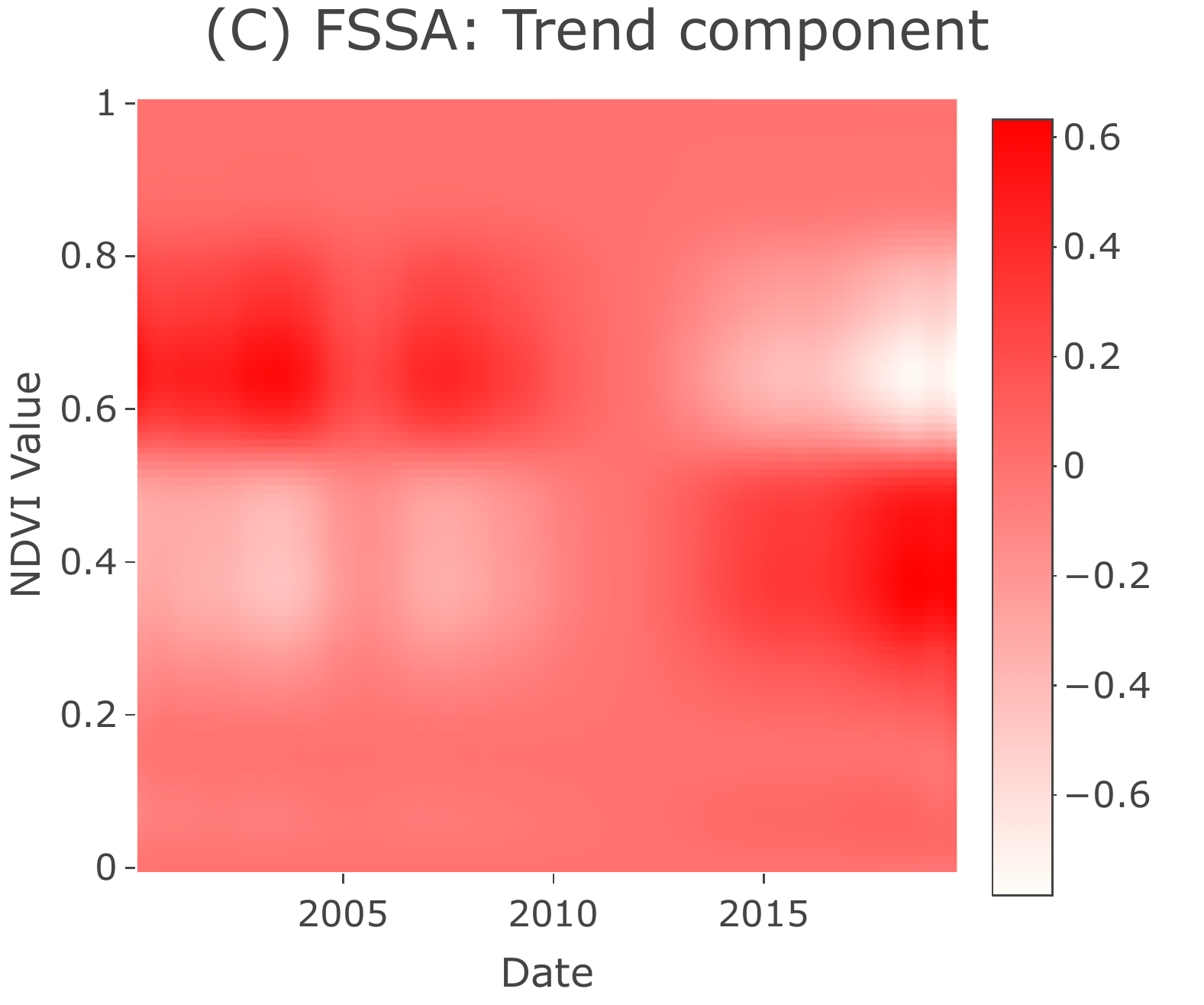}
	\end{subfigure}	
	\begin{subfigure}[b]{0.4\textwidth}
		\includegraphics[width=\textwidth]{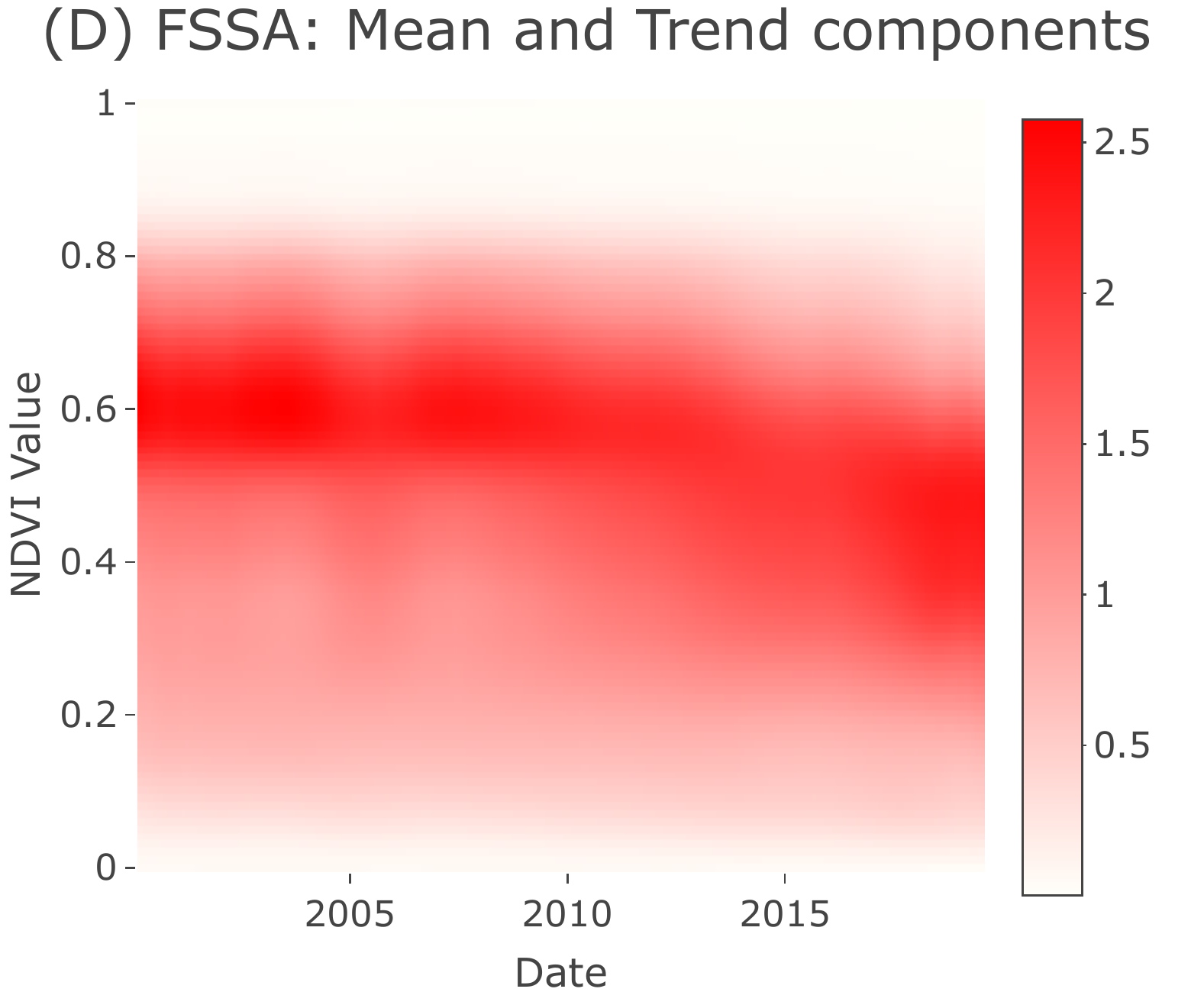}
	\end{subfigure}	
\caption{NDVI image reconstructions using (A) the overall mean component (first group); (B) annual harmonic components (second group); (C) trend component (third group); and (D) sum of the trend and overall mean components (first and third groups).}
\label{fig:ndvi_recon}
\end{center}
\end{figure}

An important goal of analyzing the NDVI data is to investigate the existence of a temporal trend in the NDVI images over time. It is interesting to see that FSSA can distinguish and separate the overall mean structure (top-left subfigures in Figures \ref{fig:ndvi_group}(D-F)) and the trend pattern (bottom-right subfigures in Figures \ref{fig:ndvi_group}(D-F)) in two different components (the first and fourth components). The trend component is causing an interesting change-point behavior after almost a decade (bottom-right subfigures in Figures \ref{fig:ndvi_group}(D-F) and Figure \ref{fig:ndvi_recon}(C)). We would like to report that SSA algorithm with the lag, $L=45$, cannot separate this trend structure, and the overall mean pattern. Therefore, SSA combines these two (the overall mean and the trend components) into one component (the SSA results are omitted due to space constraints). 

Next, we build the reconstruction of the NDVI images using the grouping suggested by the FSSA exploratory plots, $G$, as shown in Figure \ref{fig:ndvi_recon}. It is clear that Figure \ref{fig:ndvi_recon}(A) shows the reconstructed overall mean, where it does not change over time. Figure \ref{fig:ndvi_recon}(B) provides the annual harmonic behavior, and one may observe the change-point behavior after a decade in Figure \ref{fig:ndvi_recon}(C). The last subfigure, Figure \ref{fig:ndvi_recon}(D), presents the sum of the trend and overall mean components. 

It would be of interest to confirm the properties of the reconstructed groups using some rigorous statistical procedures. In order to do this, we provide the \textit{multivariate trace} periodicity test of \citet{hormann2018testing} to test for the annual seasonal effect of FTS, and a \textit{bootstrap} procedure to test existence of a trend \citep{pastecs2018} over the time series of the mean of the coefficients associated to the B-spline basis. We obtain the results of these two tests (periodicity and trend) on four sets of FTS (original signal $y_t(s)$, $R_1$, $R_2$, $R_3$), where $R_i$ represents residual curves obtained via removing the reconstructed FTS by the first $i$ groups, from the original signal $y_t(s)$. Table \ref{tab:real-NDVI} provides the p-values of the tests for the periodicity and the trend. It is clear that periodicity test captures the annual pattern for $y_t(s)$ and $R_1$ that contain the seasonal components (p-value=0). Also the trend test is significant for all of the FTS, except $R_3$ that does not contain the third group (fourth component). 

\begin{table}[!h]
\centering
\def\arraystretch{0.65}
\begin{tabular}{lcc}
 \hline
FTS & periodicity test & trend test\\ \hline
$y_{t}(s)$ & 0.00 & 0.02 \\ 
$R_{1}$   & 0.00 & 0.03 \\ 
$R_{2}$   & 0.45 & 0.03 \\ 
$R_{3}$   & 0.30 & 0.07 \\ 
 \hline
\end{tabular}
\caption{P-values of the \textit{multivariate trace} periodicity test of \citet{hormann2018testing} and the \textit{bootstrap} trend test  \citep{pastecs2018} on four FTS ($y_t(s)$, $R_1$, $R_2$, $R_3$).}
\label{tab:real-NDVI}
\end{table}

While the FSSA procedure was able to separate out the year long seasonal pattern, it was also able to detect the less obvious trend component present in the NDVI data; which indicates a loss of vegetation over the previous 20 years. It was determined that between 2001 and 2015, grassland and shrubs in the Jambi region accounted for a large amount of the vegetation lost due to controlled burning for human use \citep{Prasetyo2016}. The region specified in images that we study here is primarily upland agriculture and grass and it appears to have been a hot spot for controlled fires \citep{Prasetyo2016}.

\section{Discussion}\label{discussion}
In this paper, we constructed the FSSA procedure by incorporating the FDA techniques in basic SSA via MFPCA. The contribution of the proposed model is to provide practitioners with some tools to utilize the advantages of SSA in FTS. Accordingly, the researchers can analyze functional sequences (e.g., time series, longitudinal, or spatial data) via FSSA. Alternatively one may approach the problem using MSSA, given that the data points are measured in fixed, regular grid points over time.

As for the ease of use, an efficient and user-friendly R implementation of FSSA is developed in the \pkg{Rfssa} package. Furthermore, a shiny web application is also included in the package, and it is available at \url{https://fssa.shinyapps.io/fssa/} for reproducing the results of this paper or analyzing any other FTS. \vspace{-.25in}

\bibliographystyle{apalike}
\bibliography{Mybib}

\begin{thebibliography}{}

\bibitem[Alexandrov, 2009]{alexandrov2008method}
Alexandrov, T. (2009).
\newblock {A method of trend extraction using singular spectrum analysis}.
\newblock {\em RevStat}, 7(1):1--22.

\bibitem[Bosq, 2000]{bosq2000}
Bosq, D. (2000).
\newblock {\em Linear processes in function spaces: Theory and applications}.
\newblock Number 149 in Lecture notes in statistics. Springer, New York.

\bibitem[Chiou et~al., 2014]{chiou2014}
Chiou, J.-M., Chen, Y.-T., and Yang, Y.-F. (2014).
\newblock {Multivariate functional principal component analysis: A
  normalization approach}.
\newblock {\em Statistica Sinica}, pages 1571--1596.

\bibitem[Christensen, 1995]{christensen1995}
Christensen, O. (1995).
\newblock Frames and pseudo-inverses.
\newblock {\em Journal of Mathematical Analysis and Applications},
  195(2):401--414.

\bibitem[Fraiman et~al., 2014]{fraiman2014detecting}
Fraiman, R., Justel, A., Liu, R., and Llop, P. (2014).
\newblock Detecting trends in time series of functional data: A study of
  antarctic climate change.
\newblock {\em Canadian Journal of Statistics}, 42(4):597--609.

\bibitem[Golyandina et~al., 2015]{rssapackage}
Golyandina, N., Korobeynikov, A., Shlemov, A., and Usevich, K. (2015).
\newblock {Multivariate and 2D extensions of singular spectrum analysis with
  the {Rssa} package}.
\newblock {\em Journal of Statistical Software}, 67(2):1--78.

\bibitem[Golyandina et~al., 2018]{golyandina2018singular}
Golyandina, N., Korobeynikov, A., and Zhigljavsky, A. (2018).
\newblock {\em {Singular spectrum analysis with R}}.
\newblock Springer.

\bibitem[Golyandina et~al., 2001]{golyandina2001analysis}
Golyandina, N., Nekrutkin, V., and Zhigljavsky, A.~A. (2001).
\newblock {\em {Analysis of time series structure: SSA and related
  techniques}}.
\newblock Chapman and Hall/CRC.

\bibitem[Golyandina and Osipov, 2007]{golyandina2007caterpillar}
Golyandina, N. and Osipov, E. (2007).
\newblock {The ``Caterpillar''-SSA method for analysis of time series with
  missing values}.
\newblock {\em Journal of Statistical Planning and Inference},
  137(8):2642--2653.

\bibitem[Golyandina and Zhigljavsky, 2013]{golyandina2013singular}
Golyandina, N. and Zhigljavsky, A. (2013).
\newblock {\em {Singular spectrum analysis for time series}}.
\newblock Springer Science \& Business Media.

\bibitem[Grosjean and Ibanez, 2018]{pastecs2018}
Grosjean, P. and Ibanez, F. (2018).
\newblock {\em pastecs: Package for analysis of space-time ecological series}.
\newblock R package version 1.3.21.

\bibitem[Haghbin et~al., 2019]{rfssapackage}
Haghbin, H., Najibi, S.~M., Trinka, J., and Maadooliat, M. (2019).
\newblock {\em {Rfssa: Functional singular spectrum analysis}}.
\newblock R package version 1.0.0.

\bibitem[Happ and Greven, 2018]{happ2016}
Happ, C. and Greven, S. (2018).
\newblock Multivariate functional principal component analysis for data
  observed on different (dimensional) domains.
\newblock {\em Journal of the American Statistical Association},
  113(522):649--659.

\bibitem[Hassani and Mahmoudvand, 2018]{hassani2018singular}
Hassani, H. and Mahmoudvand, R. (2018).
\newblock {\em {Singular spectrum analysis: Using R}}.
\newblock Springer.

\bibitem[H{\"o}rmann et~al., 2015]{hormann2015}
H{\"o}rmann, S., Kidzi{\'n}ski, {\L}., and Hallin, M. (2015).
\newblock Dynamic functional principal components.
\newblock {\em Journal of the Royal Statistical Society: Series B (Statistical
  Methodology)}, 77(2):319--348.

\bibitem[H{\"o}rmann and Kokoszka, 2012]{hormann2012}
H{\"o}rmann, S. and Kokoszka, P. (2012).
\newblock {\em Functional time series}, volume~30 of {\em Handbook of
  Statistics}, pages 157--186.
\newblock Elsevier B.V., Netherlands.

\bibitem[H{\"o}rmann et~al., 2018]{hormann2018testing}
H{\"o}rmann, S., Kokoszka, P., Nisol, G., et~al. (2018).
\newblock Testing for periodicity in functional time series.
\newblock {\em The Annals of Statistics}, 46(6A):2960--2984.

\bibitem[Huang et~al., 2008]{huang2008functional}
Huang, J.~Z., Shen, H., Buja, A., et~al. (2008).
\newblock Functional principal components analysis via penalized rank one
  approximation.
\newblock {\em Electronic Journal of Statistics}, 2:678--695.

\bibitem[Kondrashov et~al., 2010]{kondrashov2010gap}
Kondrashov, D., Shprits, Y., and Ghil, M. (2010).
\newblock {Gap filling of solar wind data by singular spectrum analysis}.
\newblock {\em Geophysical Research Letters}, 37(15).

\bibitem[Lambin, 1999]{Lambin1999}
Lambin, E.~F. (1999).
\newblock {Monitoring forest degradation in tropical regions by remote sensing:
  Some methodological issues}.
\newblock {\em Global Ecology and Biogeography}, 8(3-4):191--198.

\bibitem[Maadooliat et~al., 2015]{maadooliat2015integrating}
Maadooliat, M., Huang, J.~Z., and Hu, J. (2015).
\newblock Integrating data transformation in principal components analysis.
\newblock {\em Journal of Computational and Graphical Statistics},
  24(1):84--103.

\bibitem[Mahmoudvand and Rodrigues, 2016]{mahmoudvand2016missing}
Mahmoudvand, R. and Rodrigues, P.~C. (2016).
\newblock {Missing value imputation in time series using singular spectrum
  analysis}.
\newblock {\em International Journal of Energy and Statistics}, 4(01):1650005.

\bibitem[Mohammad and Nishida, 2011]{mohammad2011comparing}
Mohammad, Y. and Nishida, T. (2011).
\newblock {On comparing SSA-based change point discovery algorithms}.
\newblock In {\em 2011 IEEE/SICE International Symposium on System Integration
  (SII)}, pages 938--945. IEEE.

\bibitem[Moskvina and Zhigljavsky, 2003]{moskvina2003algorithm}
Moskvina, V. and Zhigljavsky, A. (2003).
\newblock {An algorithm based on singular spectrum analysis for change-point
  detection}.
\newblock {\em Communications in Statistics-Simulation and Computation},
  32(2):319--352.

\bibitem[Panuju and Trisasongko, 2012]{Panuju2012}
Panuju, D.~R. and Trisasongko, B.~H. (2012).
\newblock Seasonal pattern of vegetative cover from {NDVI} time-series.
\newblock {\em Tropical Forests}, page 255.

\bibitem[Prasetyo et~al., 2016]{Prasetyo2016}
Prasetyo, L.~B., Dharmawan, A.~H., Nasdian, F.~T., and Ramdhoni, S. (2016).
\newblock {Historical forest fire occurrence analysis in Jambi Province during
  the period of 2000--2015: Its distribution \& land cover trajectories}.
\newblock {\em Procedia Environmental Sciences}, 33:450--459.

\bibitem[Ramsay and Silverman, 2007]{ramsay2007}
Ramsay, J.~O. and Silverman, B.~W. (2007).
\newblock {\em {Applied functional data analysis: Methods and case studies}}.
\newblock Springer.

\bibitem[Rodrigues and Mahmoudvand, 2016]{rodrigues2016correlation}
Rodrigues, P.~C. and Mahmoudvand, R. (2016).
\newblock {Correlation analysis in contaminated data by singular spectrum
  analysis}.
\newblock {\em Quality and Reliability Engineering International},
  32(6):2127--2137.

\bibitem[Shen and Huang, 2005]{shen2005analysis}
Shen, H. and Huang, J.~Z. (2005).
\newblock {Analysis of call centre arrival data using singular value
  decomposition}.
\newblock {\em Applied Stochastic Models in Business and Industry},
  21(3):251--263.

\bibitem[Silverman, 1986]{Silverman1986}
Silverman, B. (1986).
\newblock {\em Density estimation for statistics and data analysis}.
\newblock Chapman \& Hall, London.

\bibitem[Simon, 1980]{simon1980methods}
Simon, B. (1980).
\newblock {\em {Methods of modern mathematical physics: Functional analysis}}.
\newblock Academic Press.

\bibitem[Tuck et~al., 2014]{Tuck2014}
Tuck, S.~L., Phillips, H.~R., Hintzen, R.~E., Scharlemann, J.~P., Purvis, A.,
  and Hudson, L.~N. (2014).
\newblock {MODISTools -- downloading and processing MODIS remotely sensed data
  in R}.
\newblock {\em Ecology and Evolution}, 4(24):4658--4668.

\bibitem[Weidmann, 1980]{weidmann1980linear}
Weidmann, J. (1980).
\newblock {\em {Linear operators in Hilbert spaces}}.
\newblock Graduate texts in mathematics. Springer-Verlag.

\bibitem[Werner, 2006]{werner2006funktionalanalysis}
Werner, D. (2006).
\newblock {\em Funktionalanalysis}.
\newblock Springer.

\end{thebibliography}

\newpage

\section*{Supplementary Materials}

\subsection*{Application to call center dataset}\label{sec:call}
To illustrate the advantages of FSSA, especially its main capability in extracting different functional components (i.e. trend, harmonic and noise), we explore the call center dataset analyzed in \cite{maadooliat2015integrating}. This dataset provides the number of calls to a call center per 6 minutes intervals, between January 1 through December 31, 1999. Suppose $y_t(s_i),\ t=1,\ldots365, \ i=1,\ldots, 240$, is the square root of number of calls during the time interval $s_i$ on day $t$. Figure \ref{fig:motivating_call} (left) shows the projection of the $\{y_t(s_i)\}$'s (vectors of length $240$) into a functional space spanned by a cubic B-spline using GCV criterion.

An important goal of analyzing the call center data is to investigate the existence of periodic behaviors (e.g., weekly or monthly). Figure \ref{fig:motivating_call} (right) visually confirms the existence of a strong weekly pattern in the dataset. Since one cannot visually confirm the presence of a monthly behavior using similar graphs, it would be interesting to show that FSSA can provide tools and machinery to extract such weaker signals.

\begin{figure}[!h]
\begin{center}
	\begin{subfigure}[b]{0.45\textwidth}
		\includegraphics[page=1,width=\textwidth]{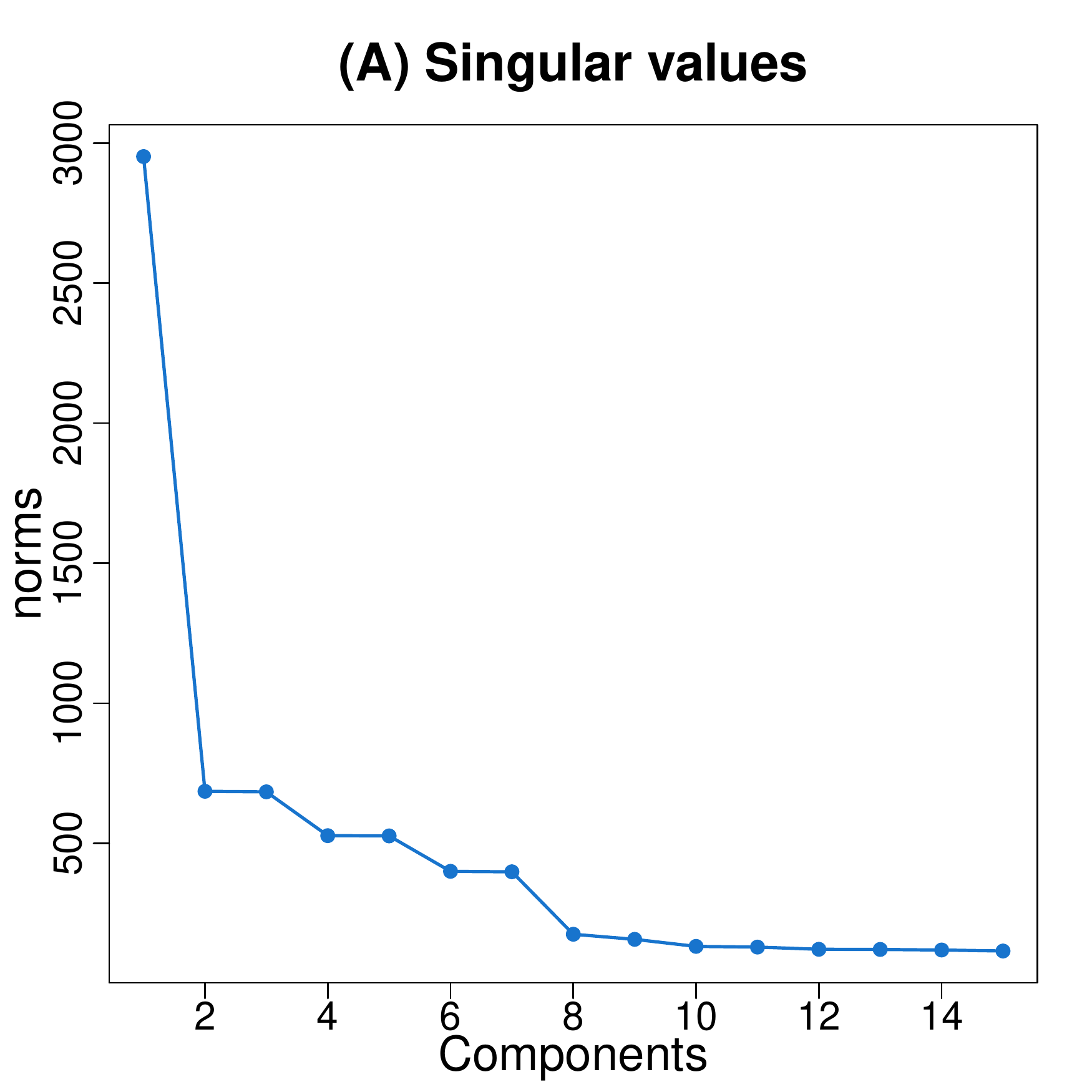}
	\end{subfigure}	
	\begin{subfigure}[b]{0.45\textwidth}
		\includegraphics[page=2,width=\textwidth]{FSSA-calls}
	\end{subfigure}	
	\begin{subfigure}[b]{0.45\textwidth}
		\includegraphics[page=3,width=\textwidth]{FSSA-calls}
	\end{subfigure}	
	\begin{subfigure}[b]{0.45\textwidth}
		\includegraphics[page=4,width=\textwidth]{FSSA-calls}
	\end{subfigure}	
\caption{FSSA plots for the grouping steps of the call center dataset.}
\label{fig:call_group}
\end{center}
\end{figure}

\begin{figure}[!h]
\begin{center}
	\begin{subfigure}[b]{0.45\textwidth}
		\includegraphics[page=5,width=\textwidth]{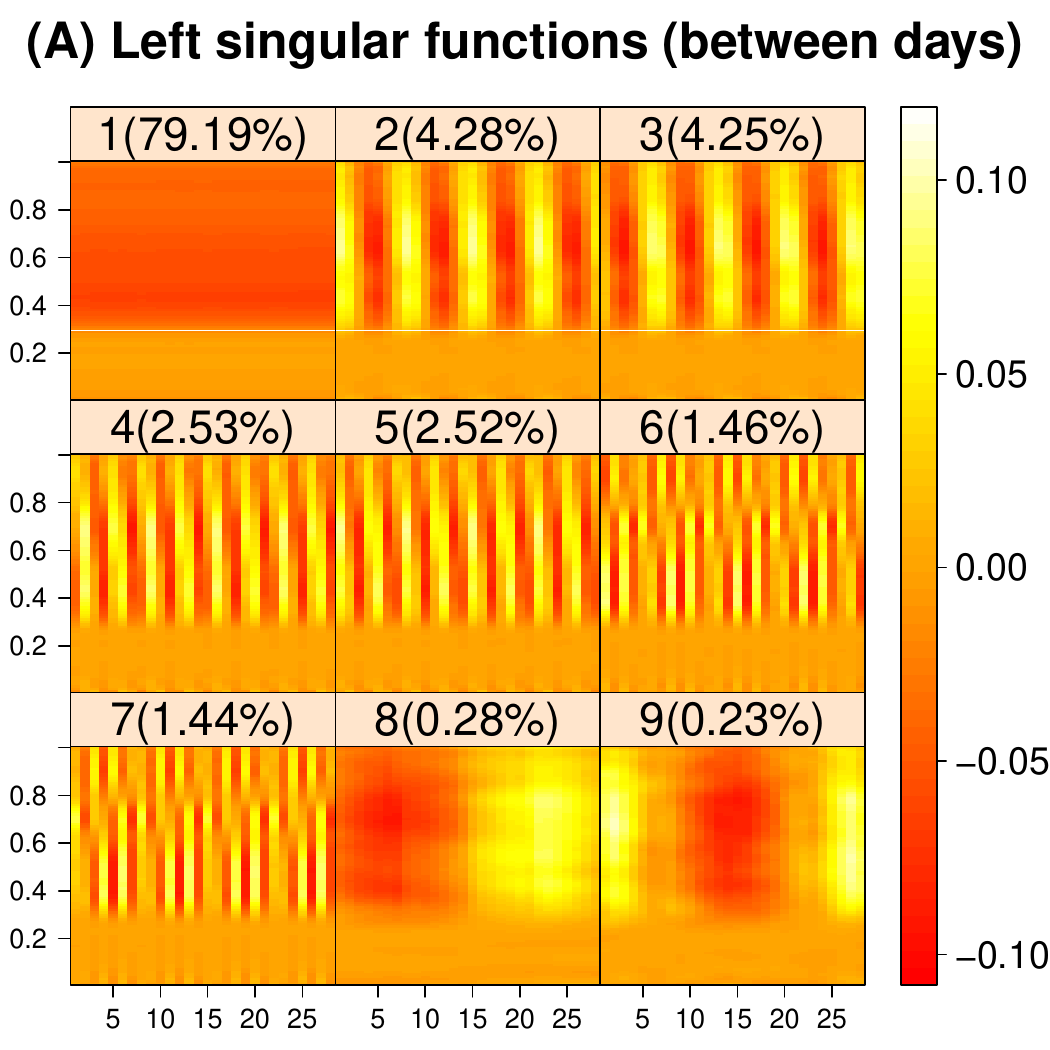}
	\end{subfigure}	
	\begin{subfigure}[b]{0.45\textwidth}
		\includegraphics[page=6,width=\textwidth]{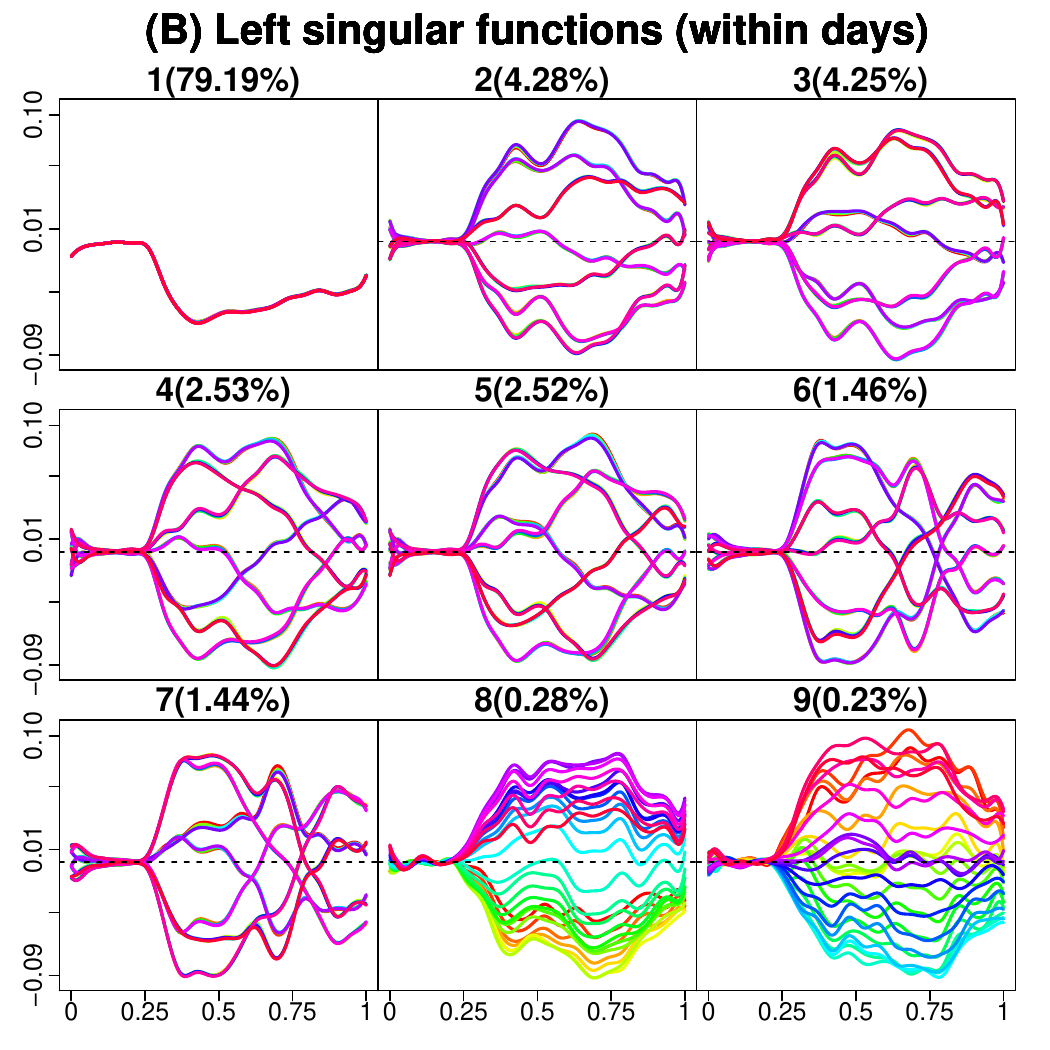}
	\end{subfigure}	
\caption{Between days (left) and within days (right) pattern of singular functions for call center dataset.}
\label{fig:call_decomp}
\end{center}
\end{figure}

In order to capture both monthly and weekly pattern by FSSA, first, we choose window length as multiple of $7$ and close to $30$, i.e., $L=28$. Then, we provide several plots using \pkg{Rfssa} package for grouping the components (Figure \ref{fig:call_group}). These plots are the functional form (analogy) of the ones commonly used in the SSA literature \citep{golyandina2001analysis}. As it can be seen in the plot of leading singular values (scree plot), the first singular value is relatively large, and there exists three evident pairs with almost equal leading singular values correspond to the three components. The w-correlation plot suggests partitioning the eigentriples into five groups: $1$, $2-3$, $4-5$, $6-7$, $8-9$ and the remainder that does not seem to contain any strong signal. Considering the remaining plots (right singular vectors and pairs of singular vectors), one can see the eigentriple pairs $2-3$, $4-5$ and $6-7$ are related to a one-week periodicity with frequencies $1/7$, $2/7$ and $3/7$, while the last group, eigentriple pair $8-9$, describes a weak monthly cycle. These groups can reproduce the reconstructed FTS. The functional components associated with the first four groups are presented in Figure \ref{fig:call_recons} (bottom) from left to right. Furthermore, some creative visualization tools that are implemented in the \pkg{Rfssa} package can be used to extract within/between days patterns for the call center data by employing the estimated left singular functions (Figure \ref{fig:call_decomp}). It is worth to mention that in Figure \ref{fig:call_decomp} (right), there are 28 curves associated with all nine singular functions (graphs). One may note that for the first singular function, all curves resemble a similar pattern respective to the main trend. Furthermore, singular functions $2-7$ contain seven distinguish patterns that each consists of four curves whereas $28$ distinct curves construct singular functions $8-9$.

For further clarification, we provide the \textit{multivariate trace} periodicity test of \citet{hormann2018testing} on six sets of FTS (original signal $y_t(s)$, $R_1$, $R_2$, $R_3$, $R_4$, $R_5$), where $R_i$ represents residual curves obtained via removing the reconstructed FTS by the first $i$ groups, from the original signal $y_t(s)$. Table \ref{tab:real} provides the p-values of the test for the periods of length $7$ and $30$ days (p-values for testing the weekly and monthly patterns). It is clear that periodicity test captures the weekly pattern for $y_t(s)$, $R_1$, $R_2$ and $R_3$ that contain either all or part of the weekly components (p-value=0). After subtracting the functional mean of all curves (first eigentriple) and the weekly components (eigentriple 2-7), the monthly pattern in $R_4$ is not weak anymore, and the periodicity test can capture the monthly cycle in $R_4$ (p-value=0). Finally, $R_5$ is the remainder of the signal after removing all of the weekly and monthly components, and that's why the associated p-values in the last row are not significant anymore.

\begin{table}[!h]
\centering
\def\arraystretch{0.65}
\begin{tabular}{lrr}
 \hline
FTS & d=7 & d=30 \\ 
 \hline
 $y_t(s)$ & 0 & 0.86 \\ 
 $R_1$ & 0 & 0.96 \\ 
 $R_2$ & 0 & 0.97 \\ 
 $R_3$ & 0 & 0.30 \\ 
 $R_4$ & 1 & 0.00 \\ 
 $R_5$ & 1 & 0.17 \\ 
 \hline
\end{tabular}
\caption{P-values of the \textit{multivariate trace} periodicity test of \citet{hormann2018testing} on six sets of FTS ($y_t(s)$, $R_1$, $R_2$, $R_3$, $R_4$, $R_5$) for the periods of length $7$ and $30$ days.}
\label{tab:real}
\end{table}

\subsection*{Proof of theorems and propositions}

\begin{proof}[Proof of Lemma]\ref{lemma:lse}
\begin{small}
\begin{align*}
\sum_{i=1}^N\Vert x_i-y\Vert^2
&=
\sum_{i=1}^N\Vert x_i-\bar{x}\Vert^2
+
2\sum_{i=1}\langle x_i-\bar{x}, \bar{x}-y \rangle
+N\Vert \bar{x}-y\Vert^2\\
&=
\sum_{i=1}^N\Vert x_i-\bar{x}\Vert^2
+N\Vert \bar{x}-y\Vert^2\geq 
\sum_{i=1}^N\Vert x_i-\bar{x}\Vert^2.
\end{align*}
\end{small}
\end{proof}

\begin{proof}[Proof of Prop]\ref{prop:trajO}
For given ${\pmb{a}}\in\mathbb{R}^K$ and ${\pmb z}\in \mathbb{H}^L$ we have
\begin{align*}
\langle \mathbfcal{X}{\pmb{a}}, {\pmb z}\rangle_{\mathbb{H}^L}&=
\sum_{j=1}^K a_j \langle {\pmb x}_j, {\pmb z}\rangle_{\mathbb{H}^L}
=\sum_{i=1}^L\sum_{j=1}^K a_j \langle y_{i+j-1}, z_i \rangle \\
&= \left\langle 
\begin{pmatrix}
a_1\\
\vdots\\
a_K\end{pmatrix},
\begin{pmatrix}
\sum_{i=1}^L \langle y_i, z_i\rangle\\
 \vdots\\
 \sum_{i=1}^L \langle y_{i+K-1}, z_i\rangle
\end{pmatrix}
\right\rangle_{\mathbb{R}^K}
=\langle {\pmb{a}}, \mathbfcal{X}^*{\pmb z}\rangle_{\mathbb{R}^K}.
\end{align*}
\end{proof}

\begin{proof}[Proof of Prop]\ref{prop_compact}
The linearity follows by bilinear form of the inner product. To prove self-adjoint property, for given ${\pmb z, w}\in \mathbb{H}^L$, \eqref{eq: s-operator} gives 
\begin{align*}
\langle\boldsymbol{\mathcal{S}}{\pmb z} , {\pmb w} \rangle_{\mathbb{H}^L}
=\sum_{j=1}^K \langle {\pmb x}_j , {\pmb z} \rangle_{\mathbb{H}^L} 
 \langle{\pmb x}_j , {\pmb w} \rangle_{\mathbb{H}^L}
=\langle{\pmb z} , \mathbfcal{S}{\pmb w} \rangle_{\mathbb{H}^L},
\end{align*}
which implies $\boldsymbol{\mathcal{S}}$ is self-adjoint. Moreover, we have 
\begin{align*}
\langle\boldsymbol{\mathcal{S}}{\pmb z} , {\pmb z} \rangle_{\mathbb{H}^L}
=\sum_{j=1}^K |\langle {\pmb x}_j , {\pmb z} \rangle_{\mathbb{H}^L}|^2\geq0, 
\end{align*}
which implies that $\boldsymbol{\mathcal{S}}$ is positive definite. To prove boundedness, let
$C=\sum_{j=1}^K \Vert {\pmb x}_j\Vert_{\mathbb{H}^L}^2\geq0$. Then
\begin{equation*}
\Vert \boldsymbol{\mathcal{S}}{\pmb z}\Vert_{\mathbb{H}^L}
=
\Vert \sum_{j=1}^K \langle {\pmb x}_j , {\pmb z} \rangle_{\mathbb{H}^L} {\pmb x}_j \Vert_{\mathbb{H}^L} 
\leq 
\sum_{j=1}^K \vert \langle {\pmb x}_j , {\pmb z} \rangle_{\mathbb{H}^L}\vert \Vert {\pmb x}_j \Vert_{\mathbb{H}^L}
\leq 
C \Vert {\pmb z} \Vert. 
\end{equation*}
As an immediate result of boundedness one can show the continuity. If ${\pmb z}\in \mathbb{H}^L$ and
 ${\pmb z}_n$ is a sequence in $\mathbb{H}^L$ such that
 $\Vert {\pmb z}_n-{\pmb z} \Vert_{\mathbb{H}^L} \rightarrow 0$ as $n\rightarrow \infty$,
then we have 
\begin{equation*}
\Vert \boldsymbol{\mathcal{S}}{\pmb z}_n-\boldsymbol{\mathcal{S}}{\pmb z} 
\Vert_{\mathbb{H}^L}
\leq C \Vert {\pmb z}_n-{\pmb z} \Vert_{\mathbb{H}^L} \rightarrow 0.
\end{equation*}
In other words, $\boldsymbol{\mathcal{S}}{\pmb z}_n\rightarrow \boldsymbol{\mathcal{S}}{\pmb z}$, which proves the continuity of $\boldsymbol{\mathcal{S}}$. To prove compactness, define $B:=\max_{i=1,\ldots, N} \Vert { y}_{i}\Vert^2$, and let $\{{\pmb f}_{n}\}$ be a weak-null sequence in $\mathbb{H}^L$, i.e.
\begin{equation*}
\langle {\pmb f}_{n}, {\pmb a} \rangle_{\mathbb{H}^L}\rightarrow 0 \ as\ n\rightarrow \infty,
\end{equation*} 
for all $\pmb{a}\in {\mathbb{H}^L}$. Then using \eqref{kernel1} and \eqref{s-oper}, with some efforts, we have
\begin{align*}
\Vert {\boldsymbol{\mathcal{S}}{\pmb f}_n}\Vert_{\mathbb{H}^L}^2&
=\sum_{j=1}^L\Vert \sum_{i=1}^L \sum_{k=1}^K y_{j+k-1}\langle f_{n,i} , y_{i+k-1}\rangle \Vert^2
=\sum_{j=1}^L\Vert \sum_{k=1}^K y_{j+k-1} \langle {\pmb f}_n , {\pmb x}_{k}\rangle_{\mathbb{H}^L} \Vert^2 \\
&\leq \sum_{j=1}^L \left( \sum_{k=1}^K\Vert y_{j+k-1} \Vert^2 \right) \left( \sum_{k=1}^K \langle {\pmb f}_n , {\pmb x}_{k}\rangle_{\mathbb{H}^L}^2\right) \leq LKB \sum_{k=1}^K \langle {\pmb f}_n , {\pmb x}_{k}\rangle_{\mathbb{H}^L}^2.
\end{align*}
Therefore, $\boldsymbol{\mathcal{S}}{\pmb f}_n \rightarrow 0$ as $n\rightarrow 0$, which implies 
 the compactness (e.g. \cite{weidmann1980linear} Thm 6.3).
\end{proof}

\begin{proof}[Proof of Prop]\ref{prop:rank1O}
For any $i\in \mathbb{N}$, define 
$\textbf{s}_i:=\bigl( \langle \pmb{\psi}_i, \pmb{x}_1\rangle_{\mathbb{H}^L},\langle \pmb{\psi}_i, \pmb{x}_2\rangle_{\mathbb{H}^L},\allowbreak \ldots, \langle \pmb{\psi}_i, \pmb{x}_K\rangle_{\mathbb{H}^L} \bigr)^\top$. Then using the definition of the operator $\mathbfcal{X}_i$, for a given ${\pmb{a}}\in\mathbb{R}^K$, we have 
 \begin{equation*}
\mathbfcal{X}_i \pmb{a}=\langle\pmb{\psi}_i,\sum_{j=1}^K a_j{\pmb x}_j\rangle_{\mathbb{H}^L} \pmb{\psi}_i
=\sum_{j=1}^K a_j \langle\pmb{\psi}_i, {\pmb x}_j\rangle_{\mathbb{H}^L} \pmb{\psi}_i 
 =\langle \textbf{s}_i,\pmb{a} \rangle_{\mathbb{R}^K}\pmb{\psi}_i,
\end{equation*}
which yields that $\mathbfcal{X}_i$ is a bounded operator of rank one (e.g. \cite{weidmann1980linear} Thm. 6.1). Moreover, since $\left\lbrace \pmb{\psi}_i, \ i\in \mathbb{N}\right\rbrace$ is an orthonormal basis system for $\mathbb{H}^L$, for all $\pmb{a}\in\mathbb{R}^K$ we have
\begin{align*}
\sum_{i=1}^\infty \mathbfcal{X}_i\pmb{a}&=\sum_{i=1}^\infty \sum_{j=1}^K a_j (\pmb{\psi}_i\otimes \pmb{\psi}_i){\pmb x}_j
=\sum_{j=1}^K a_j\sum_{i=1}^\infty \langle {\pmb{\psi}_i, \pmb x}_j \rangle_{\mathbb{H}^L} \pmb{\psi}_i\\
&=\sum_{j=1}^K a_j {\pmb x}_j =\mathbfcal{X}\pmb{a}.
\end{align*}
\end{proof}

\begin{proof}[Proof of Thm]\ref{thm:svd}
Note that $\boldsymbol{\mathcal{S}}^\dag:\mathbb{R}^K\rightarrow\mathbb{R}^K$ while $\boldsymbol{\mathcal{S}}:\mathbb{H}^L\rightarrow\mathbb{H}^L.$ Furthermore,
\begin{eqnarray*}
\boldsymbol{\mathcal{S}}^\dag \pmb{v}_l &=
\begin{pmatrix}
\sum_{j=1}^K\sum_{i=1}^L\langle y_{i}, y_{i+j-1}\rangle\frac{\langle\pmb{\psi}_l, {\pmb x}_j\rangle_{\mathbb{H}^L}}{\sqrt{\lambda_l}} \\
\vdots \\
\sum_{j=1}^K\sum_{i=1}^L\langle y_{i+K-1}, y_{i+j-1}\rangle\frac{\langle\pmb{\psi}_l, {\pmb x}_j\rangle_{\mathbb{H}^L}}{\sqrt{\lambda_l}}
\end{pmatrix}
=\frac{1}{\sqrt{\lambda_l}}
\begin{pmatrix}
\sum_{j=1}^K\langle {\pmb x}_j, {\pmb x}_1\rangle_{\mathbb{H}^L}\langle\pmb{\psi}_l, {\pmb x}_j\rangle_{\mathbb{H}^L} \\
\vdots \\
\sum_{j=1}^K\langle {\pmb x}_j, {\pmb x}_K\rangle_{\mathbb{H}^L}\langle\pmb{\psi}_l, {\pmb x}_j\rangle_{\mathbb{H}^L}
\end{pmatrix}\\
&=\frac{1}{\sqrt{\lambda_l}}
\begin{pmatrix}
\langle \sum_{j=1}^K  \langle\pmb{\psi}_l, {\pmb x}_j\rangle_{\mathbb{H}^L} {\pmb x}_j, {\pmb x}_1\rangle_{\mathbb{H}^L}\\
\vdots \\
\langle \sum_{j=1}^K  \langle\pmb{\psi}_l, {\pmb x}_j\rangle_{\mathbb{H}^L}{\pmb x}_j, {\pmb x}_K\rangle_{\mathbb{H}^L}
\end{pmatrix}
=\frac{1}{\sqrt{\lambda_l}}
\begin{pmatrix}
\langle \boldsymbol{\mathcal{S}} \pmb{\psi}_l , {\pmb x}_1\rangle_{\mathbb{H}^L}\\
\vdots \\
\langle \boldsymbol{\mathcal{S}} \pmb{\psi}_l , {\pmb x}_K\rangle_{\mathbb{H}^L}
\end{pmatrix}
=\lambda_l \pmb{v}_l .
\end{eqnarray*}
To prove orthonormality, we have
\begin{eqnarray*}
\langle \pmb{v}_l , \pmb{v}_k \rangle_{\mathbb{R}^K}
&=\frac{1}{\sqrt{\lambda_l\lambda_k}}
\sum_{j=1}^K \langle \pmb{\psi}_l , {\pmb x}_j\rangle_{\mathbb{H}^L}
 \langle \pmb{\psi}_k , {\pmb x}_j\rangle_{\mathbb{H}^L}
=\frac{1}{\sqrt{\lambda_l\lambda_k}}
\langle \pmb{\psi}_l ,\sum_{j=1}^K \langle \pmb{\psi}_k , {\pmb x}_j\rangle_{\mathbb{H}^L}{\pmb x}_j\rangle_{\mathbb{H}^L}\\
&=\frac{1}{\sqrt{\lambda_l\lambda_k}}
\langle \pmb{\psi}_l ,\boldsymbol{\mathcal{S}} \pmb{\psi}_k\rangle_{\mathbb{H}^L}
=\sqrt{\frac{\lambda_k}{\lambda_l}}
\langle \pmb{\psi}_l , \pmb{\psi}_k\rangle_{\mathbb{H}^L}=\delta_{l,k}.
 \end{eqnarray*}
 Moreover, using \eqref{eq:traj} we have
\begin{eqnarray*}
	\mathbfcal{X}{\pmb v}_i =\frac{1}{\sqrt{\lambda_i}}\sum_{j=1}^K \langle \pmb{\psi}_i , {\pmb x}_j\rangle_{\mathbb{H}^L}{\pmb x}_j
 =\frac{1}{\sqrt{\lambda_i}} \boldsymbol{\mathcal{S}} \pmb{\psi}_i  
=\sqrt{\lambda_i}\pmb{\psi}_i .
\end{eqnarray*}
\end{proof}

\begin{proof}[Proof of Thm]\ref{thm:fSep}
The Equation \eqref{winn} gives
\begin{footnotesize}
\begin{align*}
\left\langle \textbf{y}_N^{(1)} , \textbf{y}_N^{(2)} \right\rangle_w&=\sum_{i=1}^N w_i \langle y_i^{(1)} , y_i^{(2)}\rangle\\
&= \sum_{i=1}^{L-1} i \langle y_i^{(1)} , y_i^{(2)}\rangle+\sum_{i=L}^{K} L \langle y_i^{(1)} , y_i^{(2)}\rangle+\sum_{i=K+1}^{N} (N-i+1) \langle y_i^{(1)} , y_i^{(2)}\rangle\\
&=\sum_{s=2}^{L} (s-1) \langle y_{s-1}^{(1)} , y_{s-1}^{(2)}\rangle+\sum_{s=L+1}^{K+1} L \langle y_{s-1}^{(1)} , y_{s-1}^{(2)}\rangle+\sum_{i=K+2}^{N+1} (N-s+2) \langle y_{s-1}^{(1)} , y_{s-1}^{(2)}\rangle\\
&=\sum_{s=2}^{L} \sum_{j=1}^{s-1} \langle y_{s-1}^{(1)} , y_{s-1}^{(2)}\rangle+\sum_{s=L+1}^{K+1} \sum_{j=s-L}^{s-1} \langle y_{s-1}^{(1)} , y_{s-1}^{(2)}\rangle+\sum_{i=K+2}^{K+L} \sum_{j=s-L}^{K} \langle y_{s-1}^{(1)} , y_{s-1}^{(2)}\rangle\\
&=\sum_{k=1}^{K} \sum_{s=k+1}^{L+k} \langle y_{s-1}^{(1)} , y_{s-1}^{(2)}\rangle=\sum_{k=1}^{K} \sum_{i=1}^{L} \langle y_{i+k-1}^{(1)} , y_{i+k-1}^{(2)}\rangle=\sum_{k=1}^{K} \langle {\pmb x}_{k}^{(1)} , {\pmb x}_{k}^{(2)}\rangle_{\mathbb{H}^L}.
\end{align*}
\end{footnotesize}
Hence, separability of $\textbf{y}_N^{(1)}$ and $\textbf{y}_N^{(2)}$ implies $\langle {\pmb x}_{k}^{(1)} , {\pmb x}_{k}^{(2)}\rangle_{\mathbb{H}^L}=0$ for all $k=1,\ldots, K$, which complete the proof. 
\end{proof}

\begin{proof}[Proof of Lemma]\ref{basis-lemma}
The proof will be divided into two steps. In the first step it will be shown that 
 $\mathbb{H}_d^L=sp\{{\pmb \phi}_{1}, \ldots, {\pmb \phi}_{Ld}\}$. In the second step it will be proved that ${\pmb \phi}_{1}, \ldots, {\pmb \phi}_{Ld}$ are linearly independent. Suppose 
that ${\pmb z}= \left(z_1, \ldots, z_L\right)^\top$, where $z_i\in \mathbb{H}_d$. By definition, each elements of ${\pmb z}$ admits the basis expansions
$z_j=\sum_{i=1}^d b_{i,j}\nu_i,\ j=1,\ldots, L$. Therefore
\begin{small}
\begin{align}
{\pmb z}&=
\begin{pmatrix}
z_1\\z_2\\ \vdots \\ z_L
\end{pmatrix}
=
\begin{pmatrix}
\sum_{i=1}^d b_{i,1}\nu_i(s)\\
\sum_{i=1}^d b_{i,2}\nu_i(s)\\
 \vdots \\ 
\sum_{i=1}^d b_{i,L}\nu_i(s)
\end{pmatrix}
=b_{1,1}
\begin{pmatrix}
\nu_1\\0\\ \vdots \\ 0
\end{pmatrix}+ \ldots+
b_{1,L}
\begin{pmatrix}
0\\ \vdots \\ 0\\ \nu_1
\end{pmatrix}+
b_{2,1}
\begin{pmatrix}
\nu_2\\0\\ \vdots \\ 0
\end{pmatrix}\notag \\
& \qquad\qquad + \ldots+
b_{2,L}\begin{pmatrix}
0\\ \vdots \\ 0\\ \nu_2
\end{pmatrix}
+\ldots+
b_{d,1}
\begin{pmatrix}
\nu_d\\0\\ \vdots \\ 0
\end{pmatrix}+ \ldots+
b_{d,L}
\begin{pmatrix}
0\\ \vdots \\ 0\\ \nu_d
\end{pmatrix}=\sum_{k=1}^{Ld} b_{q_k,r_k} {\pmb \phi}_{k},
\end{align}
\end{small}
which implies the first step. To prove linear independency, if $\sum_{k=1}^{Ld} a_k {\pmb \phi}_{k}={\pmb 0}$ then,
\begin{small}
\begin{align*}
\begin{pmatrix}
0\\
0\\ 
\vdots \\
 0
\end{pmatrix}
&=a_1
\begin{pmatrix}
\nu_1\\0\\ \vdots \\ 0
\end{pmatrix}+ \ldots+
a_L
\begin{pmatrix}
0\\ \vdots \\ 0\\ \nu_1
\end{pmatrix} +
a_{L+1}
\begin{pmatrix}
\nu_2\\0\\ \vdots \\ 0
\end{pmatrix}+ \ldots+
a_{2L}
\begin{pmatrix}
0\\ \vdots \\ 0\\ \nu_2
\end{pmatrix}\notag \\ 
&\qquad\qquad\qquad\qquad\qquad\qquad\qquad+\ldots+
a_{(d-1)L+1}
\begin{pmatrix}
\nu_d\\0\\ \vdots \\ 0
\end{pmatrix}+ \ldots+
a_{dL}
\begin{pmatrix}
0\\ \vdots \\ 0\\ \nu_d
\end{pmatrix} \notag
\end{align*}
\end{small}

\begin{align*}
&= \begin{pmatrix}
a_1\nu_1+ a_{L+1}\nu_2+\ldots+a_{(d-1)L+1}\nu_d\\
a_2\nu_1+ a_{L+2}\nu_2+\ldots+a_{(d-1)L+2}\nu_d\\ 
\vdots \\
 a_L\nu_1+ a_{2L}\nu_2+\ldots+a_{dL}\nu_d\\ 
\end{pmatrix}
= \begin{pmatrix}
\sum_{j=1}^{d} a_{(j-1)L+1}\nu_j\\
\sum_{j=1}^{d} a_{(j-1)L+2}\nu_j\\ 
\vdots \\
 \sum_{j=1}^{d} a_{jL}\nu_j
\end{pmatrix}.
\end{align*}
This means $\sum_{j=1}^{d} a_{(j-1)L+i}\nu_j=0$ for all $i=1, \ldots, L$ and consequently $a_{(j-1)L+i}=0$ for $j=1, \ldots, d$, since $\{\nu_i\}_{i=1}^d$ are linear independent. 
\end{proof}

\begin{proof}[Proof of Thm]\ref{s-elementThm}
The proof (i) is clear, and (iii) is straightforward from (ii). 
To prove the part (ii), using \eqref{kernel1} gives
 \begin{align*}
\notag
\left\langle \mathbfcal{S}{\pmb \phi}_{i} , {\pmb \phi}_{j}\right\rangle_{\mathbb{H}^L}&=\int_0^1\int_0^1 c_{r_i,r_j}(s,u)\nu_{q_i}(s)\nu_{q_j}(u)ds du\\
&=\sum_{m=1}^K\int_0^1\int_0^1 y_{r_i+m-1}(s)y_{r_j+m-1}(u)\nu_{q_i}(s)\nu_{q_j}(u)ds du \notag\\
&=\sum_{m=1}^K\langle y_{r_i+m-1} , \nu_{q_i}\rangle \langle y_{r_j+m-1} , \nu_{q_j}\rangle.
\end{align*}
Finally, to prove (iv), the Equation \eqref{A} gives,
\begin{equation}
\label{gram}
 \begin{pmatrix}
		\langle {\pmb z}, {\pmb \phi}_{1} \rangle_{\mathbb{H}^L}\\
		\langle {\pmb z}, {\pmb \phi}_{2} \rangle_{\mathbb{H}^L}\\
 		\vdots\\
 		 \langle {\pmb z}, {\pmb \phi}_{Ld} \rangle_{\mathbb{H}^L}
 \end{pmatrix}=
 \begin{bmatrix}
		\langle {\pmb \phi}_{1}, {\pmb \phi}_{1} \rangle_{\mathbb{H}^L}& \ldots & \langle {\pmb \phi}_{Ld}, 						{\pmb \phi}_{1} \rangle_{\mathbb{H}^L}\\
		\langle {\pmb \phi}_{1}, {\pmb \phi}_{2} \rangle_{\mathbb{H}^L}& \ldots & \langle {\pmb \phi}_{Ld}, 						{\pmb \phi}_{2} \rangle_{\mathbb{H}^L}\\
		\vdots & \ddots & \vdots\\
		\langle {\pmb \phi}_{1}, {\pmb \phi}_{Ld} \rangle_{\mathbb{H}^L}& \ldots & \langle {\pmb \phi}_{Ld}, 						{\pmb \phi}_{Ld} \rangle_{\mathbb{H}^L}\\
 \end{bmatrix}
 \begin{pmatrix}
		\langle {\pmb z}, \widetilde{{\pmb\phi}}_{1} \rangle_{\mathbb{H}^L}\\
		\langle {\pmb z}, \widetilde{{\pmb\phi}}_{2} \rangle_{\mathbb{H}^L}\\
 		\vdots\\
 		 \langle {\pmb z}, \widetilde{{\pmb\phi}}_{Ld} \rangle_{\mathbb{H}^L}
 \end{pmatrix}.
 \end{equation}
 The Gram matrix $\mathbf{G}$ is Hermitian and nonsingular, since the basis ${\pmb \phi}_{i}$ are linearly independent. Let $\mathbf{G}^{-1}:=\left[h_{i,j}\right]_{i,j=1}^{Ld}$ be the inverse of $\mathbf{G}$. The Equation \eqref{gram} gives
 \begin{equation*}
 \langle {\pmb z}, \widetilde{{\pmb\phi}}_{j} \rangle_{\mathbb{H}^L}=\sum_{k=1}^{Ld} h_{j,k} \langle {\pmb z}, {\pmb \phi}_{k} \rangle_{\mathbb{H}^L},
 \end{equation*}
 which implies
 \begin{equation*}
\widetilde{{\pmb\phi}}_{j}=\sum_{k=1}^{Ld} h_{j,k} {\pmb \phi}_{k}.
 \end{equation*}
 Applying this yields 
\begin{align}
\label{sij2}
\langle \mathbfcal{S}{\pmb \phi}_{i} , \widetilde{{\pmb\phi}}_{j}\rangle_{\mathbb{H}^L}&=
\sum_{k=1}^{Ld} h_{j,k} \left\langle \mathbfcal{S}{\pmb \phi}_{i} , {\pmb \phi}_{k}\right\rangle_{\mathbb{H}^L}.
\end{align}
which implies 
$ \mathbf{S}=\mathbf{G}^{-1}\mathbf{S}_0$. \\
\end{proof}

\subsection*{R-package for FSSA routine}
\pkg{Rfssa} package includes the code and a shiny app, \url{https://fssa.shinyapps.io/fssa/}, to perform the FSSA algorithm, and present the illustrative figures described in this work. The package also contains the NDVI remote sensing and the call center datasets used as examples in the article. (available in CRAN)

\end{document}